\documentclass[12pt]{article}
\usepackage{amssymb,epic,graphicx,amsmath,color,lscape}

\allowdisplaybreaks[1]

\newcommand{\mf}[1]{{\mathfrak{#1}}}
\newcommand{\be}{\begin{eqnarray}}
\newcommand{\ee}{\end{eqnarray}}
\newcommand{\nn}{\nonumber}
\newcommand{\al}{\alpha}
\newcommand{\de}{\delta}
\newcommand{\ga}{\gamma}
\newcommand{\Ga}{\Gamma}

\newcommand{\da}{{\dot{\alpha}}}
\newcommand{\db}{{\dot{\beta}}}

\newcommand{\eps}{\epsilon}

\newcommand{\llangle}{\left\langle}
\newcommand{\rrangle}{\right\rangle}
\newcommand{\KEz}{K(E_{10})}
\newcommand{\KEn}{K(E_{9})}
\newcommand{\ds}{\varepsilon}
\newcommand{\tr}{\text{Tr}}
\newcommand{\reals}{\mathbb{R}}
\newcommand{\Z}{\mathbb{Z}}
\newcommand{\psicomp}{\psi}
\newcommand{\psicompt}{\varphi}
\newcommand{\psione}{\eta}
\newcommand{\Ups}{h}
\newcommand{\del}{\partial}
\newcommand{\op}{$\oplus$}

\newcommand{\bone}{{\bf 1}}

\newcommand{\Kbt}[3]{{G}^{(#1)#2}{}_{#3}}

\newcommand{\Fat}[2]{{Z}^{(#1)}_{#2}}

\newcommand{\Eat}[2]{{Z}^{(#1)#2}}

\newcommand{\Xat}[1]{X^{(#1)}}

\newcommand{\Yat}[1]{Y^{(#1)}}
\newcommand{\Fza}[2]{{\stackrel{(#1)}{F}}{}_{#2}}

\newcommand{\Eza}[2]{{\stackrel{(#1)}{E}}{}^{#2}}

\newcommand{\Jzg}[2]{J_{(#1)}^{#2}}
\newcommand{\Jg}[2]{J_{(#1)}^{#2}}
\newcommand{\Jgb}[2]{{S}_{(#1)}^{#2}}

\newcommand{\Xso}[2]{X^{(#1)#2}}
\newcommand{\Yso}[2]{Y^{(#1)#2}}
\newcommand{\Aso}[2]{A^{(#1)#2}}
\newcommand{\Sso}[2]{S^{(#1)#2}}

\makeatletter

\@addtoreset{equation}{section}
\makeatother

\begin{document}

\begin{center}

\begin{flushright}AEI-2006-088\\[10mm]\end{flushright}

{\Large \bf $\KEn$ from $\KEz$}\\[20mm]

Axel Kleinschmidt, Hermann Nicolai and Jakob Palmkvist\\[2mm]

{\sl Max Planck Institute for Gravitational Physics, Albert Einstein
  Institute\\
Am M\"uhlenberg 1, 14476 Potsdam, Germany}\\[15mm]

\begin{abstract}

\footnotesize{
We analyse the M-theoretic generalisation of the tangent space structure
group after reduction of the $D=11$ supergravity
theory to two space-time dimensions in the context of hidden
Kac--Moody symmetries. The action of the resulting infinite-dimensional
`R symmetry' group $K(E_9)$ on certain unfaithful, finite-dimensional
spinor representations inherited from $K(E_{10})$ is studied. We explain
in detail how these representations are related to certain finite
codimension ideals within $K(E_9)$, which we exhibit explicitly,
and how the known, as well as new finite-dimensional `generalised
holonomy groups' arise as quotients of $K(E_9)$ by these ideals.
In terms of the loop algebra realisations of $E_9$ and $K(E_9)$ on the
fields of maximal supergravity in two space-time dimensions, these
quotients are shown to correspond to (generalised) evaluation maps,
in agreement with previous results of \cite{NiSa05}. The outstanding
question is now whether the related unfaithful representations
of $K(E_{10})$ can be understood in a similar way.
}

\end{abstract}

\end{center}

\begin{section}{Introduction}

The study of a one-dimensional bosonic geodesic $\sigma$-model based
on the the Kac--Moody coset $E_{10}/K(E_{10})$ has revealed a
tantalizing dynamical link to the bosonic dynamics of maximal $D=11$
supergravity in the vicinity of a space-like singularity
\cite{DaHeNi02} (see also \cite{DaNi04}).\footnote{An alternative
  covariant approach to the bosonic dynamics of $D=11$ supergravity
  based on $E_{11}$ and the conformal group was initiated in \cite{We00,We01}.
  See also \cite{EnHo04} for a proposal combining some features of
  \cite{We01} and \cite{DaHeNi02}.
} Though striking, this link is limited to
truncations on both the Kac--Moody side and the supergravity
side. Further progress is partly
inhibited by a lack of understanding of the structure of $E_{10}$ and
of its maximal compact subgroup $K(E_{10})$ which is not even of
Kac--Moody type \cite{KlNi05}. The extension of the partial results
in the bosonic sector to fermionic fields requires the representation
theory of the infinite-dimensional $K(E_{10})$. As an important first
step it was shown in \cite{dBHP05,DaKlNi06a,dBHP06,DaKlNi06b} that
$K(E_{10})$ admits (unfaithful) spinor representations of dimensions
$320$ and $32$ with the correct properties to parallel the promising
features of the bosonic model. In particular, it was shown there
that the fermionic field equations of maximal supergravity
(with appropriate truncations) take the form of a $\KEz$ covariant
`Dirac equation'. Furthermore, the decomposition of these
spinor representations under those subgroups of $K(E_{10})$ which are
known to lead to the massive type IIA and type IIB theories were shown
to result in precisely the right (respectively, vector-like and chiral)
fermionic field representations of type IIA and type IIB supergravity
\cite{KlNi06} (the corresponding embeddings of the bosonic sectors had
already been studied previously in \cite{KlNi04a,KlNi04b} for $E_{10}$,
and \cite{SchnWe02,SchnWe01,We04a} for $E_{11}$). In this way the $E_{10}$
and $K(E_{10})$ structures incorporate kinematically and dynamically
the known duality relations between the maximal supergravity
theories for bosons and fermions alike.\footnote{The correct
  structure for the non-maximal type I supergravity theory in $D=10$
  is $DE_{10}\subset E_{10}$ \cite{HiKl06}. The ${\bf 32}$
  and ${\bf 320}$ representations of $K(E_{10})$
  decompose into the correct spinors of $K(DE_{10})$.
  The bosonic sector of this theory was previously
  studied in relation to $DE_{11}$ in \cite{SchnWe04}.}

In this paper we extend the analysis of the unfaithful $K(E_{10})$
representations to a decomposition under its $K(E_9)$ subgroup. The
latter is the maximal compact subgroup of the affine $E_9$ which
is known to be a symmetry of the field equations of maximal $N=16$
supergravity in $D=2$ \cite{Ju80,Ni87,NiWa89,NiSa98}.\footnote{See also
\cite{Ge71,BrMa87} for similar infinite-dimensional symmetries
  in pure Einstein gravity.} While the finite-dimensional
exceptional `hidden symmetries' $E_n$ of maximal supergravity in
$D=11-n$ for $n\leq 8$ can be directly realised on the supergravity
fields \cite{CrJu78,CrJuLuPo98}, the infinite-dimensional affine
symmetries of the $D=2$ theory are realised via a linear system
whose integrability condition yields the equations of motion.
The fermionic
fields (as well as the supercharges) form linear representations of the maximal
compact subgroup $K(E_n)$ for $n\le 9$. Here we will show how, using
$K(E_{10})$ and its representations, the $K(E_9)$ transformation rules
for the fermions in two space-time dimensions can be derived purely
algebraically from the reduction. This constitutes the first direct proof of
the $K(E_9)$ properties of $D=2$ supergravity that does not resort to
the linear system. Moreover, we will show that our algebraic action is
equivalent to the analytic description of the $K(E_9)$ action in terms
of a spectral parameter via a `generalised evaluation map' \cite{NiSa05}.
The equivalence of the latter with the algebraic construction of the
present work suggests that $K(E_{10})$ may admit a similar realisation
-- a tantalizing opportunity for future research,
since it may also lead to a new realisation of the hyperbolic Kac--Moody
algebra $E_{10}$ itself!

A major tool in our investigation is the so-called level decomposition
of the global hidden symmetries $E_n$. In fig.~\ref{e10dynk}
below, we display the Dynkin diagram of $E_{10}$ with our labelling
conventions; the lower rank exceptional algebras are obtained
by removing nodes from the left.
\begin{figure}[t]
\begin{center}
\scalebox{1}{
\begin{picture}(340,60)
\put(5,-5){$1$} \put(45,-5){$2$} \put(85,-5){$3$}
\put(125,-5){$4$} \put(165,-5){$5$} \put(205,-5){$6$}
\put(245,-5){$7$} \put(285,-5){$8$} \put(325,-5){$9$}
\put(260,45){$10$} \thicklines
\multiput(10,10)(40,0){9}{\circle{10}}
\multiput(15,10)(40,0){8}{\line(1,0){30}}
\put(250,50){\circle{10}} \put(250,15){\line(0,1){30}}
\end{picture}}
\caption{\label{e10dynk}\sl Dynkin diagram of $E_{10}$ with numbering
of nodes.}
\end{center}
\end{figure}
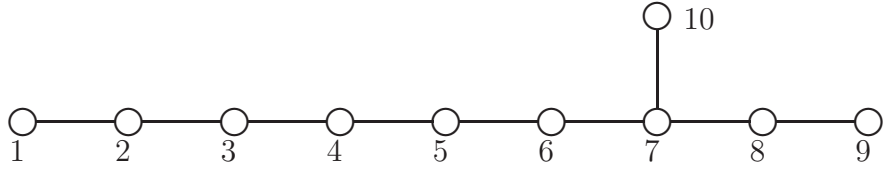
The level decomposition with regard to the $A_{n-1} \equiv \mf{sl}(n)$
subalgebras of $E_n$ allows one to identify the physical fields from
the adjoint representation of $E_n$ in terms of $SL(n)$ tensors. More
specifically, these level decompositions
follow the scheme presented in
table~\ref{levdectab} for $n=5,\ldots,9$, where we label the relevant
$SL(n)$ representations by bold face letters in the usual way, noting
that the entries of the columns $\ell=1,2$ always correspond to
the three- and six-form representations of $SL(n)$, respectively
(and the columns $\ell = -1,-2$ to their contragredient representations).
Naturally, $E_6$ in five dimensions is the first time the
six-forms appear in the scalar coset.

\begin{table}[h!]
\begin{center}
\begin{tabular}{|c|p{3.5mm}p{1.5mm}p{3.5mm}p{1.5mm}p{3.5mm}p{1.5mm}p{3.5mm}p{1mm}p{16mm}p{1.5mm}p{3.5mm}p{1.5mm}p{3.5mm}p{1.5mm}p{3.5mm}p{1.5mm}p{3.5mm}|}\hline
$E_n\backslash \ell$
&$-4$&&$-3$&&$-2$&&$-1$&&\centering0&&1&&2&&3&&4\\
\hline\hline
$E_5$ & & & & &  & & \bf{10} & \op&\centering{(\bf{24}\op\bf{1})} &\op& \bf{10} &&&&&&\\
$E_6$ & & & & & \bf{1} & \op & \bf{20} & \op &\centering (\bf{35}\op\bf{1}) & \op &
    \bf{20} &\op & \bf{1} &&&&\\
$E_7$ & & & & & \bf{7} & \op & \bf{35} & \op &\centering (\bf{48}\op\bf{1}) & \op &
     \bf{35} & \op & \bf{7} &&&&\\
$E_8$ & & &  $\overline{\bf{8}}$ & \op & \bf{28} & \op &
$\overline{\bf{56}}$ & \op &\centering (\bf{63}\op\bf{1}) & \op & \bf{56} & \op &
$\overline{\bf{28}}$ & \op & \bf{8} &&\\
$E_9$ & $\cdots$ & \op & \bf{80} & \op & \bf{84} & \op & $\overline{\bf{84}}$
  & \op &{\centering{({\bf{80}\op 1\op 1})}} & \op & \bf{84} & \op & $\overline{\bf{84}}$ &
   \op & \bf{80} & \op &
  $\cdots$ \\ \hline
\end{tabular}
\caption{\label{levdectab}\sl The level decompositions of the global $E_n$
  hidden symmetries in $D=11-n$ dimensions under the gravity $SL(n)$
  subgroup. The column headings $\ell$ refer to the level in this
  level decomposition. For $\ell=0$, the adjoint of $SL(n)$ always
  combines with the singlet into the adjoint of $GL(n)$, in the affine
  case also extended by the derivation $d$. The central element $c$ of
  $\mf{e}_9$ is part of $\mf{gl}(9)$. }
\end{center}
\end{table}

For the finite-dimensional algebras in this series (that is, for $n\leq 8$)
these results have been known for a long time (for a systematic analysis,
see \cite{CrJuLuPo98}). For $n=9$, the triple of representations
$\overline{\bf{84}}\oplus\bf{80}\oplus \bf{84}$ is repeated an infinite
number of times, giving rise to the affine extension of $E_8$ in the
standard way (the two singlets appearing in the middle column for $E_9$
are the central charge $c$ and the derivation $d$). For $n=10$ and $n=11$,
we can no longer display the
representations in such a simple fashion, as the number of representations
`explodes'; but see \cite{FiNi03} for the tables up to levels $\ell=18$
and $\ell=10$, respectively, which were obtained by computer
algebra,\footnote{In fact, for $E_{10}$, the tables are available up to
  $A_9$ level $\ell = 28$ with a total of $4\,400\,752\,653$
  representations \cite{FiNi03}.} and also \cite{DaHeNi02} and \cite{We03a} for
earlier results on very low levels of $E_{10}$ and $E_{11}$, respectively.

We conclude this introduction with some comments on the link between
the mathematical structures (ideals, and unfaithful representations
of infinite-dimensional compact subgroups of hidden symmetries) exhibited
in the main part of this paper, and the so-called 'generalised holonomies'
discussed in the recent literature. {\em Quite generally, the latter should
be identified with quotients of the infinite-dimensional algebras $\KEn$
and $\KEz$ by certain finite codimension ideals.} Given any Lie algebra
$\mf{k}$ and a linear representation space $V$, the subspace
\be\label{id}
\mf{i}_V := \big\{ x\in\mf{k} \, | \, x\cdot v = 0 \quad \forall v\in V \big\}
\subset \mf{k}
\ee
defines an {\em ideal} in $\mf{k}$. The representation is unfaithful if
$\mf{i}_V \neq \{0\}$. The existence of
non-trivial ideals implies in particular that the Lie algebra $\mf{k}$ is
not simple. For any $\mf{i}_V$, we can define the quotient algebra
\be\label{q}
\mf{q}_V := \mf{k}/ \mf{i}_V \subset \mf{gl}(V).
\ee
The unfaithful {\em finite-dimensional} spinorial representations
of $\KEn$ and $\KEz$ discovered
in \cite{NiSa05,dBHP05,DaKlNi06a,dBHP06,DaKlNi06b} are directly related
to the Dirac- and vector (gravitino) spinors appearing in maximal
supergravities. For instance, the relevant representations for $\KEz$
are the $\bf{32}$ and the ${\bf 320}$ \cite{DaKlNi06a,dBHP06}. These
representations are inherited by $\KEn\subset \KEz$, such that the
$\bf{32}$ decomposes into two inequivalent 16-dimensional Dirac-type
representations of $\KEn$. {\em As one of our main results we are able
to present the associated ideals in $\KEn$ in complete detail},
cf. section~\ref{ke9sec}. Because a single ideal may be associated to
more than one (and sometimes infinitely many) representations, the
description of these structures in terms of ideals appears to be the most
economical way to study them.

It is perhaps worth stressing that the quotient group
$SO(16)_+\times SO(16)_-$ associated to the $\bf{16}_+\oplus \bf{16}_-$
representation of $\KEn$ is {\em not} a subgroup of $K(E_{9})$, because the
would-be $SO(16)_+\times SO(16)_-$ generators are {\em distributional
objects}, as we will explain (see also \cite{DaKlNi06b}). The latter
group has been proposed as a `generalised holonomy group' of M-theory
\cite{DuLiu03,Hu03}, generalising the $SO(9)$ Lorentz structure group
of the tangent space of the nine torus on which the $D=11$ theory was
reduced. By studying its subgroups and the branching of the ${\bf 32}$
representation under these, supersymmetric solutions can be studied
and classified \cite{DuSt91,DuLiu03,Hu03,BaDuLiWe05}. On the other hand,
it is known that neither this generalised holonomy group, nor its
extensions $SO(32)$ and $SL(32)$, can extend to symmetries of the
full equations because of global obstructions \cite{Ke04}.
In addition, the generalised holonomies proposed so far do not admit
acceptable vector-spinor representations, and as such are restricted
to the Killing spinor equation instead of the full supergravity
system (in particular, the Rarita Schwinger equation). Our results
strengthen the case for $K(E_9)$ and for $\KEz$ as the correct
generalised holonomy (and R symmetry) groups since both groups
do allow for all the required spinor representations.
Moreover, $\KEn$ is a genuine local symmetry of the reduced theory.

This article is organised as follows. Section~\ref{e9sec} summarizes some
(largely known) results on the embedding of $E_8$ and $E_9$ in a notation
adapted to the level decomposition, and goes on to derive their embedding into
$E_{10}$. Informed readers may skip the bulk of this section and proceed
directly to section~\ref{ke9sec}, where we derive the branching of the
unfaithful $K(E_{10})$ spinors under the $K(E_9)$ subalgebra. The resulting
$\KEn$ transformation rules are compared to those of the linear system in
section~\ref{ke9curr}. Using relations provided in two appendices, we establish complete agreement with
previous results of \cite{NiSa05}.

\end{section}

\begin{section}{$E_8$, $E_9$ and $E_{10}$}
\label{e9sec}

We here study the chain of embeddings $E_8\subset E_9\subset E_{10}$
in $A_7\subset A_8\subset A_9$ level decompositions and fix necessary
notation for our analysis of the spinor representations in the next
section. 
Throughout this paper, except for the appendices, we adopt the following indexing conventions for
the $SL(n)$ tensors arising in the decomposition of the algebras $E_8$, $E_9$ and $E_{10}$:
\be
E_{10} \quad \leftrightarrow& \quad a,b,\ldots &\in \{1,\dots,10\} \nn\\
E_{9} \quad \leftrightarrow& \quad \alpha,\beta,\ldots
&\in \{2,\dots,10\} \nn\\
E_8 \quad \leftrightarrow& \quad i,j,\ldots &\in \{3,\dots,10\}
\ee

\begin{subsection}{$E_8$ via $A_7$} \label{e8viaa7}

The $E_8$ subalgebra of $E_{10}$ is generated by nodes $3$ through
to $10$ of fig.~\ref{e10dynk} and can be written in terms of irreducible
tensors of its $A_7\cong \mf{sl}(8)$ subalgebra (corresponding to
nodes $3$ through to $9$). By adjoining the eigth Cartan generator,
this $\mf{sl}(8)$ subalgebra can be extended to a $\mf{gl}(8)$
subalgebra generated by
\be
G^i{}_j \quad,\quad\text{with}\quad
  \left[G^{i}{}_{j},G^k{}_l\right] = \de^k_j G^i{}_l - \de^i_l G^k{}_j,
\ee
where the indices take values $i,j=3,\ldots, 10$. The $A_7$ decomposition
of $E_8$ gives the $\mf{sl}(8)$ tensors displayed in
table~\ref{tabe8a7} \cite{CrJuLuPo98}.

\begin{table}[h!]
\renewcommand{\arraystretch}{1.2}
\begin{center}
\begin{tabular}{|c|c|r|}
\hline
$A_7$ level $\ell$ in $E_8$ & Generator & $SL(8)$ representation\\
\hline
\hline
$-3$ & $Z_{i}$ & $\overline{\bf{8}}$\\
$-2$ & $Z_{i_1 ... i_6}$ & $\bf{28}$\\
$-1$ & $Z_{i_1i_2i_3}$ & $\overline{\bf{56}}$\\
$0$ & $G^i{}_j$ & ${\bf 63} \oplus {\bf 1}$\\
$1$ & $Z^{i_1i_2i_3}$ & ${\bf 56}$\\
$2$ & $Z^{i_1\ldots i_6}$ & $\overline{\bf{28}}$\\
$3$ & $Z^{i}$ & $\bf{8}$\\
\hline
\end{tabular}
\end{center}
\caption{\label{tabe8a7}\sl $A_7$ decomposition of $E_8$.}
\end{table}

In the left column we have indicated the $\mf{sl}(8)$ level, that is
the number of times the exceptional simple root $\alpha_{10}$ occurs
in the associated roots. All indices $i,j,\ldots$ run from
$3,\ldots,10$ and all tensors, except for $G^i{}_j$, are totally
anti-symmetric in their  $SL(8)$ (co-)vector indices. The Chevalley
transposition $(\cdot)^T$ acts by $(G^i{}_j)^T = G^j{}_i$ and
$(Z_{i_1i_2i_3})^T = Z^{i_1i_2i_3}$, etc. The $\mf{gl}(8)$ tensors
in the table with upper (lower) indices correspond to positive
(negative) roots. In $E_{10}$ language, the former correspond to the
`E-type' generators, while the latter transform in the contragredient
representations and correspond to the `F-type' generators in the
notation of \cite{DaNi04}.

The commutation relations between $G^i{}_j$ and the positive and
negative $\mf{gl}(8)$ level `step operators' are
\be\label{ke8}
\left[G^i{}_j , Z^{k_1k_2k_3}\right] &=&
  3 \de_j^{[k_1} Z^{k_2k_3]i},\nn\\
\left[G^i{}_j , Z^{k_1\ldots k_6}\right] &=&
  -6 \de_j^{[k_1} Z^{k_2\ldots k_6]i},\nn\\
\left[G^i{}_j , Z^{k}\right] &=&
  \de_j^k Z^{i} + \de_j^i Z^{k},\nn\\
\left[G^i{}_j , Z_{k_1k_2k_3}\right] &=&
  -3 \de^i_{[k_1} Z_{k_2k_3]j},\nn\\
\left[G^i{}_j , Z_{k_1\ldots k_6}\right] &=&
  6 \de^i_{[k_1} Z_{k_2\ldots k_6]j},\nn\\
\left[G^i{}_j , Z_{k}\right] &=&
  -\de^i_k Z_{j} - \de_j^i Z_{k}.
\ee
Note the trace terms in the commutation relations involving the
$\mf{gl}(8)$ vectors $Z^{i}$ and $Z_{i}$ which are needed for
the correct transformation under the trace of $\mf{gl}(8)$, and for
the consistency of the first two relations with the second relation
in (\ref{ee8}) below. Furthermore,
\be\label{ee8}
\left[Z^{i_1i_2i_3}, Z^{i_4i_5i_6}\right] &=&
  Z^{i_1\ldots i_6},\nn\\
\left[Z^{i_1i_2i_3}, Z^{i_4\ldots i_9}\right] &=&
  3 Z^{[i_1}\eps^{i_2i_3]i_4\ldots i_9},
\ee
where $\eps^{i_1\ldots i_8}$ is the $SL(8)$
totally anti-symmetric tensor. 
Similar expressions are obtained for the negative level generators
by applying the Chevalley transposition.

The mixed commutation relations are
\be\label{ef8}
\left[Z^{i_1i_2i_3} , Z_{j_1j_2j_3} \right] &=&
   -2 \de^{i_1i_2i_3}_{j_1j_2j_3} G
    + 18 \de^{[i_1i_2}_{[j_1j_2} G^{i_3]}_{\ j_3]},\nn\\
\left[ Z^{i_1i_2i_3} , Z_{j_1\ldots j_6} \right] &=&
  -5!\, \de^{i_1i_2i_3}_{[j_1j_2j_3} Z_{j_4j_5j_6]},\nn\\
\left[ Z^{i_1\ldots i_6}, Z_{j_1\ldots j_6} \right] &=&
   -\frac23\cdot 6!\, \de^{i_1\ldots i_6}_{j_1\ldots j_6} G
   + 6\cdot 6!\, \de^{[i_1\ldots i_5}_{[j_1\ldots j_5}
       G^{i_6]}_{\ j_6]},\nn\\
\left[ Z^{i_1i_2i_3} , Z_{j} \right] &=&
  \frac1{5!}\eps^{i_1i_2i_3k_1\ldots k_5} Z_{k_1\ldots k_5
    j},\nn\\
\left[ Z^{i_1\ldots i_6} , Z_j \right] &=&
  \frac12 \eps^{i_1\ldots i_6k_1k_2} Z_{k_1k_2j},\nn\\
\left[ Z^i , Z_{j} \right] &=& G^i{}_j.
\ee
Here, $ G \equiv \sum_{k=3}^{10} G^k{}_k$.
Equations (\ref{ke8}), (\ref{ee8}) and (\ref{ef8}), together with
their Chevalley transposes, constitute a complete set of $E_8$
commutation relations. The normalisations of the generators are
\be\label{E8norm}
\llangle G^i{}_j \big| G^k{}_l\rrangle &=&
  \de^i_l \de^k_j + \de^i_j\de^k_l,\nn\\
\llangle Z^{i_1i_2i_3}\big| Z_{j_1j_2j_3}\rrangle &=&
  3!\, \de^{i_1i_2i_3}_{j_1j_2j_3},\nn\\
\llangle Z^{i_1\ldots i_6}\big| Z_{j_1\ldots j_6}\rrangle &=&
  6!\, \de^{i_1\ldots i_6}_{j_1\ldots j_6},\nn\\
\llangle Z^i \big| Z_j\rrangle &=&
  \de^i_j.
\ee
Modulo normalisation factors, the same relations have been given for
example in \cite{CrJuLuPo98,KoNiSa00}. In comparison with the notation
of \cite{KoNiSa00} the tensors on levels $\ell=\pm 2$ have been dualised
using the $\eps$-tensor of $SL(8)$ and some of the normalisations
have changed.

\end{subsection}

\begin{subsection}{$E_9$ as extended current algebra}

As is well known (see e.g. \cite{Ka90}), the affine Lie algebra
$E_9\equiv E_8^{(1)}$ is
obtained from $E_8$ by `affinization', that is by embedding $E_8$ in
its current algebra (parametrized by the spectral parameter $t$), and
by adjoining two more Lie algebra elements, the central charge $c$ and
the derivation $d$: $E_9=E_8[[t,t^{-1}]]\oplus \reals c \oplus\reals d$
(as always, we restrict attention to the split
real forms of these Lie algebras).
By writing $\Xat{m} \equiv X\otimes t^m$
(for $m\in\Z$) the $E_9$ commutation relations are
\be\label{E9aff}
\big[\Xat{m}, \Yat{n}\big] &=& \big[X\otimes t^m, Y\otimes t^n\big]
  = \left[X,Y\right]\otimes t^{m+n}
   + m\de_{m+n,0} \langle X|Y \rangle\, c,\nn\\
\left[d,\Xat{m}\right] &=& m\Xat{m},\nn\\
\left[c,\Xat{m}\right] &=& 0,\quad\quad \left[c,d\right] = 0.
\ee
They can thus be read off directly from the $E_8$ commutation relations
above in the standard fashion. The inner product between $c$ and
$d$ is $\langle c|d\rangle=1$. The `horizontal' $E_8$ at affine level
$0$ is isomorphic to $E_8$ and we will often write $X\equiv\Xat{0}$
for any $E_8$ generator $X$, for example
\be
G^i{}_j \equiv \Kbt{0}{i}{j} \;,\; Z_i \equiv \Fat{0}{i} \;,\;\;\; etc.
\ee
Next, we will study how the current algebra generators emerge from
$E_{10}$, that is how they are obtained from the latter algebra by
truncation and by `dimensional reduction'.

\end{subsection}

\begin{subsection}{Embedding of $E_9$ in $E_{10}$}
\label{e9e10sec}

With regard to the $E_{10}$ Dynkin diagram, the $E_9$ subalgebra of
$E_{10}$ is obtained by deleting node $1$ from the diagram~\ref{e10dynk},
or equivalently by restricting
to level zero in an $E_9$ level decomposition\footnote{In comparison
  to the $A_9$ level decomposition of $E_{10}$ which can be thought of
  as a space-like foliation of the Lorentzian root lattice, the $E_9$
  decomposition foliates the root lattice by {\em null} planes.}
which counts the
number of occurrences of the simple root $\alpha_1$. However, one does keep
the Cartan generator $h_1$ which is needed to `desingularize' the metric
on the root lattice (so the Cartan subalgebra of $E_9$ can be
identified with the one of $E_{10}$, $h_1$ appears only in $d$). Using
the $\mf{gl}(10)$ basis of $E_{10}$, where small Latin indices take
values $a=1,\ldots,10$,
\be
K^a{}_b\quad,\quad\text{with}\quad
  \left[K^a{}_b,K^c{}_d\right] = \de^c_b K^a{}_d - \de^a_d K^c{}_b,
\ee
the $E_{10}$ Cartan generators are
\be
h_a &=& K^a{}_a - K^{a+1}{}_{a+1}\quad\quad(a=1,\ldots,9),\nn\\
\label{h10}
h_{10} &=& -\frac13\sum_{a=1}^{10}K^a{}_a
    +K^8{}_8+K^9{}_9+K^{10}{}_{10}.
\ee
The invariant inner product of these generators is given by
\be\label{E10norm}
\langle K^a{}_b|K^c{}_d\rangle = \de^a_d \de^c_b - \de^a_b\de^c_d.
\ee
The coefficient of the second term is not fixed by invariance but by
requiring that $\langle h_{10}|h_{10}\rangle=2$, where $h_{10}$ in
(\ref{h10})  was fixed by requiring the right $\mf{gl}(10)$
commutation relation with
the $A_9$ level $\ell=1$ generator $E^{abc}$.\footnote{This is the
    reason for the minus sign in the bilinear form (\ref{E10norm}),
    resulting in the indefiniteness of the inner product (\ref{E10norm}).
    By contrast, (\ref{E8norm}) has a plus sign in the corresponding
    formula, whence the inner product is positive definite for $E_8$.}
We follow the conventions of \cite{DaNi04} except for two differences:
Firstly, we take $e_{10}$ to be $E^{8\,9\,10}$  since the exceptional
node is attached at the other end. Secondly, we rescale the $A_9$
level $\ell=\pm 3$ generators by a factor $1/3$.

In terms of the $A_9$ level decomposition of $E_{10}$ the $E_9$
elements are precisely those contained in the `gradient
representations' of \cite{DaHeNi02} where indices are restricted
to take values $2,\ldots,10$. As shown there (see also \cite{FiNi03}),
every $n$th order gradient generator contains $n$ sets of nine
anti-symmetric indices, and thus has $A_9$ Dynkin labels $[n********]$.
For instance, at $A_9$ level $\ell=3n+1$, we have the following gradient
generators
\be
E^{a_1^{(1)}\cdots a_9^{(1)}|\cdots|a_1^{(n)}\cdots a_9^{(n)}|bcd}
\qquad \mbox{with $a_i^{(j)},b,c,d\in\{1,\ldots , 10\}$}\,, \nn
\ee
which are antisymmetric in all {\em 9-tuples} $(a^{(j)}_1\cdots
a^{(j)}_9)$. Restricting all indices on the above element to
the values $2,\ldots,10$, we see that, up to permutations, there
is only one choice of filling indices into these {\it 9-tuples},
and we thus only need to remember that there were $n$ such sets.
In fact, this restriction is physically motivated since $E_9$
arises in the reduction to two dimensions with one left-over
non-compact spatial direction $1$ (obviously, there are
nine alternative choices for this residual spatial dimension, corresponding
to ten distinguished $E_9$ subgroups in $E_{10}$). Accordingly, we
introduce the following shorthand notation for the gradient generators
\be\label{gradlev}
E^{\scriptsize\overbrace{2...10|\cdots|2...10|}^{n\,\text{times}}\al_1\al_2\al_3}
&\equiv& \Eza{n}{\al_1\al_2\al_3}     \nn\\
E^{\scriptsize\overbrace{2...10|\cdots|2...10|}^{n\,\text{times}}
   \al_1\dots\al_6}
&\equiv& \Eza{n}{\al_1\dots\al_6}     \nn\\
E^{\scriptsize\overbrace{2...10|\cdots|2...10|}^{n\,\text{times}}
   \al_0|\al_1\dots\al_8}
&\equiv& \Eza{n}{\al_0|\al_1\dots\al_8} \,
\ee
where $\al_0,\al_1,\al_2,\dots =2,\ldots,10$. The 'F-type' gradient
generators are defined analogously. Our notation is summarized
in table~\ref{gradtab}: the indices here take values
$\al=2,\ldots,10$, and together with $K^\al{}_\beta$ and
$K^1{}_1$ from $A_9$ level $\ell=0$ they constitute {\em all} $E_9$
generators expressed in $E_{10}$ variables. As will be seen below, the
central charge $c$ of  $E_9$ in terms of $E_{10}$ generators is
proportional to $K^1{}_1$ and commutes with all elements of $E_9$
whence the restriction of indices to $\al=2,\ldots,10$ is the correct
restriction to $E_9$. That the suppression of the blocks of nine
indices is justified will be shown below. Now we want to relate these
generators to the $E_9$ generators of section~\ref{e9sec}.

\begin{table}[t]
\begin{center}
\begin{tabular}{|r|c|c|}
\hline
$A_9$ level in $E_{10}$ & (Restricted) gradient generator &
$\mf{sl}(9)$ irrep \\
\hline \hline
$\ell = 3n+3$ & $\Eza{n}{\al_0|\al_1\ldots \al_8}$ & ${\bf 80}$\\
$\ell = 3n+2$ & $\Eza{n}{\al_1\ldots \al_6}$ & ${\bf \overline{84}}$\\
$\ell = 3n+1$ & $\Eza{n}{\al_1\al_2\al_3}$ & ${\bf 84}$\\\hline
$\ell = -3n-1$ & $\Fza{n}{\al_1\al_2\al_3}$ & ${\bf \overline{84}}$\\
$\ell = -3n-2$ & $\Fza{n}{\al_1\ldots \al_6}$ & ${\bf 84}$\\
$\ell = - 3n-3$ & $\Fza{n}{\al_0|\al_1\ldots \al_8}$  & ${\bf 80}$ \\
\hline
\end{tabular}
\caption{\label{gradtab}\sl Identification of the $E_9$ generators in
  terms of $E_{10}$ gradient generators.}
\end{center}
\end{table}

The generators of $E_8$ are embedded regularly in $E_{10}$ and, away
from the Cartan subalgebra, are identical to those of $E_{10}$ for
levels $|\ell|\leq 3$ if the indices are restricted to the range
$\{3,\dots,10\}$. Therefore we find immediately that
\begin{align}\label{e8e10}
\Eat{0}{i_1i_2i_3} &= \Eza{0}{i_1i_2i_3},&
  \Fat{0}{i_1i_2i_3} &=\Fza{0}{i_1i_2i_3},&\nn\\
\Eat{0}{i_1\ldots i_6} &= \Eza{0}{i_1\ldots i_6},&
  \Fat{0}{i_1\ldots i_6} &= \Fza{0}{i_1\ldots i_6},&\nn\\
 \eps^{k_1\ldots k_8} \Eat{0}{i} &= \Eza{0}{i|k_1\ldots  k_8},&
   \eps_{k_1\ldots k_8}\Fat{0}{i} &= \Fza{0}{i|k_1\ldots k_8},&
\end{align}
where the superscript on the l.h.s. denotes the affine level,
whereas the superscript on the r.h.s. denotes the
`gradient' level as explained in (\ref{gradlev}).
As a mnemonic and notational device to distinguish between these
two kinds of levels we place the superscripts slightly differently,
as evident from the preceding equation. The objects on the r.h.s. are
$GL(8)$ tensors, and we recall that, for the comparison between $E_8$
and $E_{10}$ we must restrict the indices on the $SL(10)$ tensors
appearing in the $A_9$ decomposition of $E_{10}$
to the values $i=3,\ldots, 10$. To identify the $GL(8)$ generators in
terms of the Cartan generators we note that the only difference can be
in the diagonal part of $\Kbt{0}{i}{j}$ since the off-diagonal elements
correspond to step operators. A simple calculation shows that the correct
identification between $G^i{}_j \in E_8$ and $K^i{}_j \in E_{10}$
is\footnote{One way to see the necessity of this redefinition is to compute
  $[\Eat{0}{8\,9\,10},\Fat{0}{8\,9\,10}]$ both in $E_8$ and $E_{10}$,
  and to demand that the central charge $c$ and the derivation $d$
  drop out from this commutator for $E_8$.
}
\be\label{kk10}
G^i{}_j \equiv \Kbt{0}{i}{j} = K^i{}_j +\de^i_j (c-d),
\ee
where the central element $c$ and derivation $d$ of $E_9$ in terms of the
$\mf{gl}(10)$ generators are given by
\be\label{cd10}
d=K^{2}{}_2,\quad\quad c=-K^1{}_1.
\ee
It is easy to see that $c$ indeed commutes with all elements of $E_9$
and has
inner product $+1$ with $d$. Furthermore, $d$ commutes with $E_8$ of
(\ref{e8e10}) as it should. Evidently, the affine level operator $d$
counts the number of tensor indices taking the value $2$ (with $(+1)$
for upper and $(-1)$ for lower indices). The extra terms in (\ref{kk10})
also induce the relative change in sign between (\ref{E8norm}) and
(\ref{E10norm}).

Using the relation of the general linear subalgebras (\ref{kk10}) we
can show that the blocks of nine anti-symmetric indices suppressed in
the gradient generators are not `seen' by the $\mf{gl}(8)$ generators,
as we already claimed above. Consider a generator $X^{2k_1\ldots k_8}$
which is totally anti-symmetric in its nine indices and $k\in\{3,\ldots ,10\}$.
Then
\be
\left[\Kbt{0}{i}{j} , X^{2k_1\ldots k_8}\right] =
  8\de_j^{[k_1}X^{k_2\ldots k_8]2i} - \de^i_j X^{2k_1\ldots k_8}
  = -9\de_j^{[i} X^{k_1\ldots k_8]2} = 0
\ee
by Schouten's identity; the last term in the middle expression is due to the
correction term with $d$ in (\ref{kk10}), which is thus crucial for the
vanishing of the above commutator. This confirms that we can indeed replace
each {\it 9-tuple} of indices by a label indicating the number of
such {\it 9-tuples} and assume that the {\it 9-tuples} are filled in
some fixed way by $2,\ldots,10$.

From the form of $d$ in (\ref{cd10}) we see that the number of upper
indices equal to $2$ on a positive step generators is the affine level and similarly
for negative step operators. It is not hard to identify the following
affine level $+1$ generators among the $E_{10}$ generators,
\be\label{aff1}
\Fat{1}{j} &=& K^2{}_j,\nn\\
\Fat{1}{j_1\ldots j_6} &=& \frac12 \eps_{j_1\ldots j_6k_1k_2}
  \Eza{0}{k_1k_2 2},\nn\\
\Fat{1}{j_1j_2j_3} &=& \frac1{5!}\eps_{j_1j_2j_3 k_1\ldots k_5}
   \Eza{0}{k_1\ldots k_5 2},\nn\\
\Kbt{1}{i}{j} &=& -\frac1{7!}\eps_{jk_1\ldots k_7}
  \Eza{0}{i|2k_1\ldots k_7}  -\frac1{8!}\de^i_j\eps_{k_1\ldots
      k_8}\Eza{0}{2|k_1\ldots k_8}.
\ee
This involves only generators with $A_9$ level $\ell=0,\ldots,3$ in
the $E_{10}$ decomposition. Similarly, at affine level $-1$ we have
\be\label{affm1}
\Eat{-1}{i} &=& K^i{}_2,\nn\\
\Eat{-1}{i_1\ldots i_6} &=& \frac12 \eps^{i_1\ldots i_6k_1k_2}
  \Fza{0}{k_1k_2 2},\nn\\
\Eat{-1}{i_1i_2i_3} &=& \frac1{5!}\eps^{i_1i_2i_3 k_1\ldots k_5}
   \Fza{0}{k_1\ldots k_5 2},\nn\\
\Kbt{-1}{i}{j} &=& -\frac1{7!}\eps^{ik_1\ldots k_7}
  \Fza{0}{j|2k_1\ldots k_7}  -\frac1{8!}\de^i_j\eps^{k_1\ldots
      k_8}\Fza{0}{2|k_1\ldots k_8}.
\ee
Again, the redefinition (\ref{kk10}) is crucial for the correct
$E_9$ transformation rules, e.g.
\be
\left[\Kbt{0}{i}{j}, \Fat{1}{k_1\ldots k_6}\right] &=&
  \frac12\eps_{k_1\ldots k_6l_1l_2} \left[K^i{}_j -\de^i_j d \,,\,
     \Eza{0}{l_1l_22}\right]\nn\\
&=& \frac12\eps_{k_1\ldots k_6l_1l_2}\left(2\de^{l_1}_j
  \Eza{0}{il_22}-\de^i_j\Eza{0}{l_1l_22}\right)\nn\\
&=& \frac1{2\cdot 6!}\eps_{k_1\ldots k_6l_1l_2}\left(2\de^{l_1}_j
  \eps^{il_2m_1\ldots m_6} - \de^i_j\eps^{l_1l_2m_1\ldots
    m_6}\right)\Fat{1}{m_1\ldots m_6} \nn\\
&=& 6 \de^i_{[k_1}\Fat{1}{k_2\ldots k_6]j},
\ee
in agreement with (\ref{ke8}) for affine level $+1$. We identify also
the following elements at affine level $\pm2$,
\be\label{affpm2}
\Fat{2}{j} &=& -\frac1{7!} \eps_{jk_1\ldots  k_7}
   \Eza{0}{2|2k_1\ldots k_7},\nn\\
\Eat{-2}{i} &=& -\frac1{7!} \eps^{ik_1\ldots  k_7}
   \Fza{0}{2|2k_1\ldots k_7}.
\ee
Indeed, one can check from these relations that
\be
\left[\Eat{-2}{i}, \Fat{2}{j}\right] = \Kbt{0}{i}{j}-2\de^i_j \, c
\ee
as it should be for this affine commutator. Again we see, that the
affine level is equal to the difference between the number of upper
and lower indices equalling  $2$. With relations
(\ref{e8e10}), (\ref{kk10}), (\ref{cd10}), (\ref{aff1}), (\ref{affm1})
and (\ref{affpm2}) we have identified all $E_9$ generators appearing
on $A_9$ levels $-3\le \ell\le 3$ in $E_{10}$. It should now be clear
how to obtain the higher affine levels: the scheme repeats itself after
shifting $\ell\rightarrow \ell +3$, as illustrated in figure~\ref{e9e10fig}.
As evident from these formul\ae{}, the affine level and the $A_9$ level
are `oblique' w.r.t. each other: The elements of affine level $n$ are
spread over the $A_9$ levels $3n-3\le\ell \le 3n+3$. This is also
shown in figure~\ref{e9e10fig}.

For completeness, we write the general formul\ae{} for $n>1$,
\be\label{e9e10a}
\Fat{n}{i} &=&
  -\frac1{7!}\eps_{ik_1\ldots k_7}\Eza{n-2}{2|2k_1\ldots k_7},\nn\\
\Fat{n}{i_1\ldots i_6} &=&
  \frac12\eps_{i_1\ldots i_6k_1k_2}\Eza{n-1}{k_1k_22},\nn\\
\Fat{n}{i_1i_2i_3} &=&
   \frac1{5!}\eps_{i_1i_2i_3k_1\ldots k_5}\Eza{n-1}{k_1\ldots
     k_52},\nn\\
G^{(n)i}{}_j &=& -\frac1{7!}\eps_{jk_1\ldots
   k_7}\Eza{n-1}{i|2k_1\ldots k_7} -\frac1{8!}\de^i_j \eps_{k_1\ldots
   k_8} \Eza{n-1}{2|k_1\ldots k_8},\nn\\
\Eat{n}{i_1i_2i_3} &=&  \Eza{n}{i_1i_2i_3},\nn\\
\Eat{n}{i_1\ldots i_6} &=& \Eza{n}{i_1\ldots i_6},\nn\\
\Eat{n}{i} &=&  \frac1{8!} \eps_{k_1\ldots k_8}
  \Eza{n}{i|k_1\ldots k_8}
\ee
for the positive current modes and
\be\label{e9e10b}
\Fat{-n}{i} &=&
  \frac1{8!}\eps^{k_1\ldots k_8}\Fza{n}{i|k_1\ldots k_8},\nn\\
\Fat{-n}{i_1\ldots i_6} &=&
  \Fza{n}{i_1\ldots i_6},\nn\\
\Fat{-n}{i_1i_2i_3} &=&   \Fza{n}{i_1i_2i_3},\nn\\
G^{(-n)i}{}_j &=& -\frac1{7!}\eps^{ik_1\ldots
   k_7}\Fza{n-1}{\!\!\!j|2k_1\ldots k_7} -\frac1{8!}\de^i_j \eps_{k_1\ldots
   k_8} \Fza{n-1}{\!\!\!2|k_1\ldots k_8},\nn\\
\Eat{-n}{i_1i_2i_3} &=&  \frac1{5!}\eps^{i_1i_2i_3k_1\ldots k_5}
   \Fza{n-1}{\!\!\!k_1\ldots k_52},\nn\\
\Eat{-n}{i_1\ldots i_6} &=& \frac12\eps^{i_1\ldots i_6k_1k_2}
   \Fza{n-1}{\!\!\!k_1k_22},\nn\\
\Eat{-n}{i} &=& -\frac1{7!} \eps^{ik_1\ldots k_7}
  \Fza{n-2}{\!\!\!2|2k_1\ldots k_7}
\ee
for the negative current modes with $n>1$. Observe that the $\mf{sl}(8)$
representations appearing in the vertical lines in fig.~\ref{e9e10fig}
combine `sideways' into the required $\mf{sl}(9)$ representations in
accordance with the decompositions
\be
{\bf{80}}  &\rightarrow& {\bf{8}} \oplus ({\bf{63}}\oplus {\bf 1}) \oplus \overline{\bf{8}},
                   \nn\\
{\bf{84}}  &\rightarrow&  {\bf{56}} \oplus {\bf{28}}, \nn\\
\overline{\bf{84}}  &\rightarrow&  \overline{\bf{56}} \oplus
        \overline{\bf{28}}.
\ee

\begin{landscape}
\begin{figure}
\input{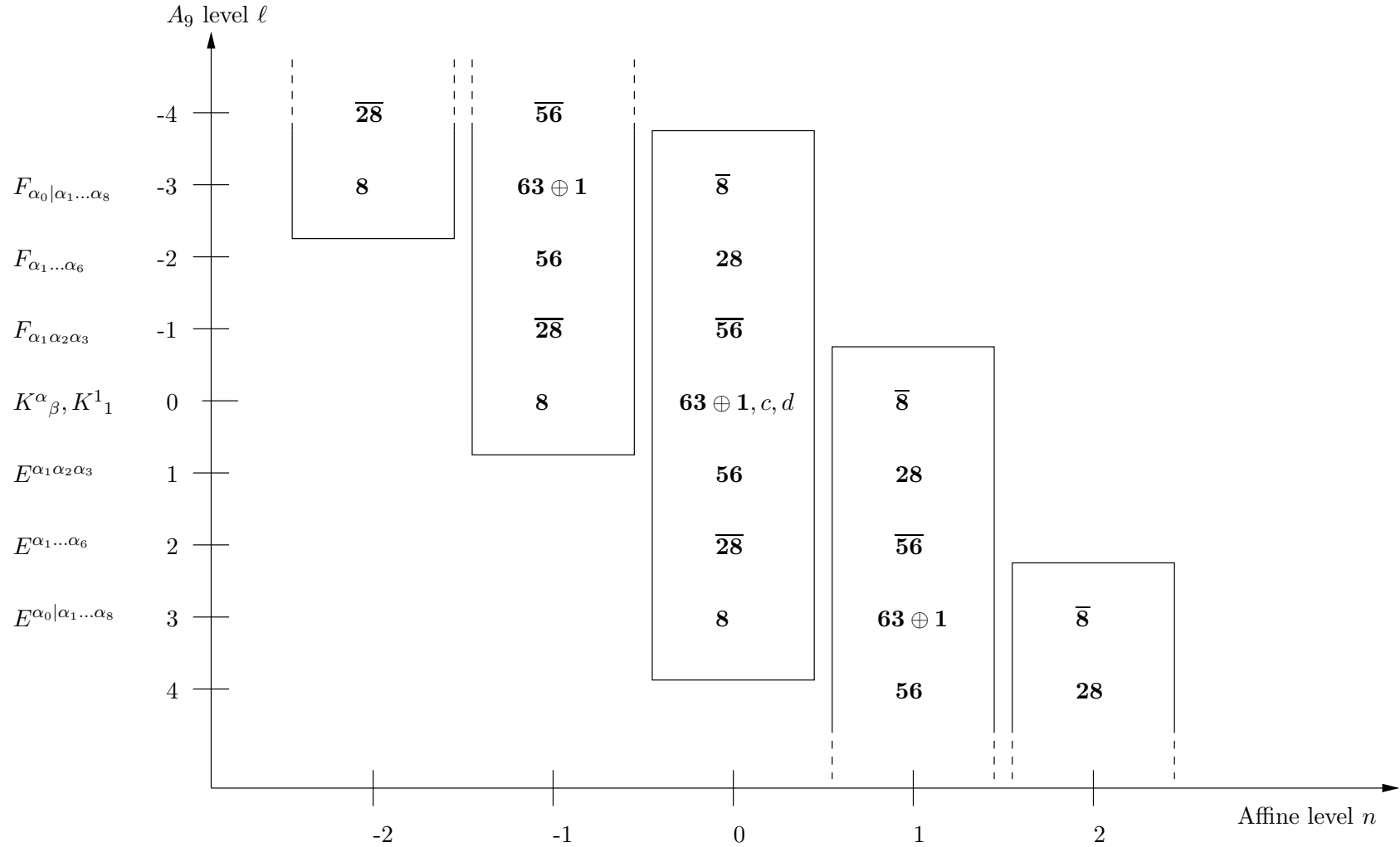}
\caption{\label{e9e10fig}\sl Diagram illustrating the distribution of
  $A_9$ levels and affine levels in $E_{10}$. The affine level $n$ is
  given by the number of upper $2$s minus the number of lower $2$s on
  an $E_{10}$ generator. The indices on the $A_9$ level $\ell\ne 0$
  generators range over $\al=2,\ldots,10$. The boxed set of generators
  correspond to copies of $E_8$, at affine level $0$, the central
  charge and derivation are also included.}
\end{figure}
\end{landscape}

\end{subsection}

\end{section}

\begin{section}{$\KEn$ spinor representations from $\KEz$}
\label{ke9sec}

The generators of $\KEz$ are the anti-symmetric elements
under the Chevalley transposition (see
e.g. \cite{DaKlNi06b}). Therefore, we can construct a $\KEz$ generators
for any positive root step operator $E$ by taking $J=E-E^T\equiv E-F$. The
restriction to $\KEn$ is then obtained by considering only those
positive step operators of table~\ref{gradtab}. As mentioned in the
introduction $\KEn$ is not of Kac--Moody type (nor is $\KEz$). The
reason for this is that the invariant inner product
\be\label{Hilbert}
(x|y) := - \llangle x | y \rrangle  \qquad \mbox{for all $x,y\in\KEn$
  (or $\KEz$)},
\ee
inherited from the invariant bilinear form on $E_9$ ($E_{10}$), is
positive definite on the compact subalgebras \cite{KlNi05}.

Despite this complication, finite-dimensional, hence {\em unfaithful},
representations corresponding to Dirac-spinor and vector-spinor
(gravitino) representations of $\KEz$ have been constructed in
\cite{dBHP05,DaKlNi06a,dBHP06,DaKlNi06b}. We now study the branching
of these representations to $\KEn\subset  \KEz$. Before doing so we
derive the complete $\KEn$ commutation relations in a form convenient
for this computation.

The $\KEz$ generators at `$A_9$ levels' $\ell=0,\ldots,3$ are defined
by
\be
\Jzg{0}{ab} &=& K^a{}_b - K^b{}_a,\nn\\
\Jzg{1}{a_1a_2a_3} &=& \Eza{0}{a_1a_2a_3} - \Fza{0}{a_1a_2a_3},\nn\\
\Jzg{2}{a_1\ldots a_6} &=& \Eza{0}{a_1\ldots a_6} - \Fza{0}{a_1\ldots
  a_6},\nn\\
\Jzg{3}{a_0|a_1\ldots a_8} &=& \Eza{0}{a_0|a_1\ldots a_8} -
  \Fza{0}{a_0|a_1\ldots a_8},
\ee
for $a,b=1,\ldots,10$. Observe that on the l.h.s. the position of
indices no longer matters, as these tensors transform only under
the $SO(10)$ subgroup of $\KEz$ and indices can be raised and lowered
with the invariant $\de^{ab}$. The lower indices in
parentheses on the l.h.s. indicate the $A_9$ level in $E_{10}$ (or
$A_8$ level in $E_9$), where as the indices placed above the generators
on the r.h.s. indicate the gradient level of (\ref{gradlev}).
As before, the $\KEn$ generators are obtained from these by `dimensional
reduction', that is by restricting the indices to $\alpha,\beta=2,\ldots,10$,
corresponding to the $A_8$ level decomposition of $E_9$. The relation
between the $A_8$ decomposition and the current algebra decomposition
of $E_9$ was explained in the preceding section.

In the remainder we will make use of the following notation for the
$\KEn$ generators in $\KEz$ for $k\ge 0$,
\be \label{ke9kompakt}
\Jg{0}{\al\beta} &=& K^\al{}_\beta - K^\beta{}_\al,\nn\\
\Jg{3k+1}{\al_1\al_2\al_3} &=&
  \Eza{k}{\al_1\al_2\al_3}-\Fza{k}{\al_1\al_2\al_3},\nn\\
\Jg{3k+2}{\al_1\ldots\al_6} &=&
  \Eza{k}{\al_1\ldots\al_6}-\Fza{k}{\al_1\ldots\al_6},\nn\\
\Jg{3k+3}{\al_0|\al_1\ldots\al_8} &=&
  \Eza{k}{\al_0|\al_1\ldots\al_8}-\Fza{k}{\al_0|\al_1\ldots\al_8},
\ee
using the notation of (\ref{gradlev}) and table~\ref{gradtab}. The
generator at $A_8$ level $(3k+3)$ is $\mf{so}(9)$ reducible and
decomposes after dualisation into
\be\label{elldg}
\Jg{3k+3}{\beta|\al_1\ldots\al_8} = \left(\Jg{3k+3}{\beta\ga} +
  \Jgb{3k+3}{\beta\ga}\right)\eps^{\ga\al_1\ldots \al_8}
   \qquad \mbox{for $k\geq 0$}.
\ee
Here, the anti-symmetric tensor $\Jg{3k+3}{\al\beta}= -\Jg{3k+3}{\beta\al}$
is the trace part of the orginal tensor $\Jg{3k+3}{\al_0|\al_1\ldots \al_8}$,
and the symmetric $\Jgb{3k+3}{\al\beta}= + \Jgb{3k+3}{\beta\al}$ is
traceless, $\Jgb{3k+3}{\ga\ga}=0$, according to the original Young
symmetry. The anti-symmetric part has the same representation
structure as $\Jg{0}{\al\beta}$; by contrast, the symmetric generators
$\Jgb{3n}{\al\beta}$ have no zero mode part, and exist only for $n\geq 1$.

{}From (\ref{e9e10a}) and (\ref{e9e10b}) we deduce the following $\KEn$
relations (for $m\ge n$),
\be\label{ke9rels}
\left[\Jg{3m}{\al\beta},\Jg{3n}{\ga\de}\right] &=&
  2\de^{\beta\ga}\Jg{3(m+n)}{\al\de}
  + 2\de^{\beta\ga}\Jg{3(m-n)}{\al\de},\nn\\
\left[\Jg{3m}{\al\beta},\Jgb{3n}{\ga\de}\right] &=&
    2\de^{\beta\ga}\Jgb{3(m+n)}{\al\de}
   - 2\de^{\beta\ga}\Jgb{3(m-n)}{\al\de},\nn\\
\left[\Jgb{3m}{\al\beta},\Jg{3n}{\ga\de}\right] &=&
    2\de^{\beta\ga}\Jgb{3(m+n)}{\al\de}
   + 2\de^{\beta\ga}\Jgb{3(m-n)}{\al\de},\nn\\
\left[\Jgb{3m}{\al\beta},\Jgb{3n}{\ga\de}\right] &=&
    2\de^{\beta\ga}\Jg{3(m+n)}{\al\de}
   - 2\de^{\beta\ga}\Jg{3(m-n)}{\al\de},\nn\\
\left[\Jg{3m}{\al\beta},\Jg{3n+1}{\ga_1\ga_2\ga_3}\right] &=&
  3\de^{\beta\ga_1}\Jg{3(m+n)+1}{\al\ga_2\ga_3}
  -\frac3{6!}\de^{\beta\ga_1}\eps^{\al\ga_2\ga_3\de_1\ldots\de_6}
    \Jg{3(m-n)-1}{\de_1\ldots\de_6},\nn\\
\left[\Jg{3n}{\al\beta},\Jg{3m+1}{\ga_1\ga_2\ga_3}\right] &=&
  3\de^{\beta\ga_1}\Jg{3(m+n)+1}{\al\ga_2\ga_3}
  +3\de^{\beta\ga_1}\Jg{3(m-n)+1}{\al\ga_2\ga_3},\nn\\
\left[\Jgb{3m}{\al\beta},\Jg{3n+1}{\ga_1\ga_2\ga_3}\right] &=&
  3\de^{\beta\ga_1}\Jg{3(m+n)+1}{\al\ga_2\ga_3}
  +\frac3{6!}\de^{\beta\ga_1}\eps^{\al\ga_2\ga_3\de_1\ldots\de_6}
    \Jg{3(m-n)-1}{\de_1\ldots\de_6}\nn\\
&&    -\frac13\de^{\al\beta}\Jg{3(m+n)+1}{\ga_1\ga_2\ga_3}
    -\frac1{3\cdot 6!}\de^{\al\beta}\eps^{\ga_1\ga_2\ga_3\de_1\ldots\de_6}
    \Jg{3(m-n)-1}{\de_1\ldots\de_6},\nn\\
\left[\Jgb{3n}{\al\beta},\Jg{3m+1}{\ga_1\ga_2\ga_3}\right] &=&
  3\de^{\beta\ga_1}\Jg{3(m+n)+1}{\al\ga_2\ga_3}
  -3\de^{\beta\ga_1} \Jg{3(m-n)-1}{\al\ga_2\ga_3}\nn\\
&&    -\frac13\de^{\al\beta}\Jg{3(m+n)+1}{\ga_1\ga_2\ga_3}
    +\frac1{3}\de^{\al\beta}\Jg{3(m-n)-1}{\ga_1\ga_2\ga_3},\nn\\
\left[\Jg{3m}{\al\beta},\Jg{3n+2}{\ga_1\dots\ga_6}\right] &=&
  6\de^{\beta\ga_1}\Jg{3(m+n)+2}{\al\ga_2\ldots\ga_6}
  -\de^{\beta\ga_1}\eps^{\al\ga_2\ldots\ga_6\de_1\de_2\de_3}
    \Jg{3(m-n)-2}{\de_1\de_2\de_3},\nn\\
\left[\Jg{3n}{\al\beta},\Jg{3m+2}{\ga_1\dots\ga_6}\right] &=&
  6\de^{\beta\ga_1}\Jg{3(m+n)+2}{\al\ga_2\ldots\ga_6}
  +6\de^{\beta\ga_1} \Jg{3(m-n)+2}{\al\ga_2\ldots \ga_6},\nn\\
\left[\Jgb{3m}{\al\beta},\Jg{3n+2}{\ga_1\dots\ga_6}\right] &=&
  6\de^{\beta\ga_1}\Jg{3(m+n)+2}{\al\ga_2\ldots\ga_6}
  +\de^{\beta\ga_1}\eps^{\al\ga_2\ldots\ga_6\de_1\de_2\de_3}
    \Jg{3(m-n)-2}{\de_1\de_2\de_3}\nn\\
&&-\frac23\de^{\al\beta}\Jg{3(m+n)+2}{\ga_1\ldots\ga_6}
  -\frac19\de^{\al\beta}\eps^{\ga_1\ldots\ga_6\de_1\de_2\de_3}
    \Jg{3(m-n)-2}{\de_1\de_2\de_3},\nn\\
\left[\Jgb{3n}{\al\beta},\Jg{3m+2}{\ga_1\dots\ga_6}\right] &=&
  6\de^{\beta\ga_1}\Jg{3(m+n)+2}{\al\ga_2\ldots\ga_6}
  -6\de^{\beta\ga_1} \Jg{3(m-n)+2}{\al\ga_2\ldots\ga_6}\nn\\
&&-\frac23\de^{\al\beta}\Jg{3(m+n)+2}{\ga_1\ldots\ga_6}
  +\frac23\de^{\al\beta}\Jg{3(m-n)+2}{\ga_1\ldots\ga_6},\nn\\
\left[\Jg{3m+1}{\al_1\al_2\al_3},\Jg{3n+1}{\beta_1\beta_2\beta_3}\right] &=&
  \Jg{3(m+n)+2}{\al_1\al_2\al_3\beta_1\beta_2\beta_3}
  -18 \de^{\al_1\beta_1}\de^{\al_2\beta_2}\left(\Jg{3(m-n)}{\al_3\beta_3}
  +\Jgb{3(m-n)}{\al_3\beta_3}\right),\nn\\
\left[\Jg{3m+1}{\al_1\al_2\al_3},\Jg{3n+2}{\beta_1\dots\beta_6}\right] &=&
  3 \eps^{\ga\beta_1\ldots \beta_6\al_1\al_2}
    \left(\Jg{3(m+n)+3}{\al_3\ga} +\Jgb{3(m+n)+3}{\al_3\ga}\right) \nn\\
&&  +\frac16\de_{\al_1\al_2\al_3}^{\beta_1\beta_2\beta_3}
    \eps^{\beta_4\beta_5\beta_6\ga_1\ldots \ga_6}
    \Jg{3(m-n)-1}{\ga_1\ldots \ga_6},\nn\\
\left[\Jg{3n+1}{\al_1\al_2\al_3},\Jg{3m+2}{\beta_1\dots\beta_6}\right] &=&
  3 \eps^{\ga\beta_1\ldots \beta_6\al_1\al_2}
    \left(\Jg{3(m+n)+3}{\al_3\ga} +\Jgb{3(m+n)+3}{\al_3\ga}\right) \nn\\
&&  -120\de_{\al_1\al_2\al_3}^{\beta_1\beta_2\beta_3}
     \Jg{3(m-n)+1}{\beta_4\beta_5\beta_6},\nn\\
\left[\Jg{3m+2}{\al_1\ldots\al_6},\Jg{3n+2}{\beta_1\dots\beta_6}\right]&=&
  -6\cdot 6!  \de^{\al_1\beta_1}\cdots\de^{\al_5\beta_5}
    \left(\Jg{3(m-n)}{\al_6\beta_6}+\Jgb{3(m-n)}{\al_6\beta_6}\right)\nn\\
&&  -400\de^{\al_1\beta_1}\cdots\de^{\al_3\beta_3}
    \eps^{\al_4\ldots\al_6\beta_4\ldots\beta_5\ga_1\ga_2\ga_3}
     \Jg{3(m+n)+4}{\ga_1\ga_2\ga_3}\,\,,
\ee
with implicit (anti-)symmetrizations on the r.h.s. according to the
symmetries of the l.h.s. and with the understanding that
the level zero symmetric piece vanishes:
$\Jgb{0}{\al\beta}=0$. Note that in some relations a level index
become negative for $m=n$; in those cases one has to use
the formula in the next row for which this does not happen.
Let us emphasize once more that these formulas
were deduced by making use of the identifications found in the previous
section, and by exploiting the fact that the affine $E_9$ commutators
are known {\em for all levels}, whereas we have no complete
knowledge of the higher level commutation relations for $E_{10}$.
From the above commutation relations, one readily verifies that the Lie algebra
$\KEn$ indeed possesses a `filtered' structure, with
\be\label{ke9filt}
[\Jg{k}{}\,,\, \Jg{l}{} ] = \Jg{k+l}{} + \Jg{|k-l|}{}
\qquad (k,l\geq 0).
\ee

\begin{subsection}{Dirac-spinor ideal}
\label{dssec}

Under $\KEz$ the $32$-dimensional Dirac-spinor $\ds$ transforms as
follows on the first four levels \cite{dBHP05,DaKlNi06a,DaKlNi06b},
\be\label{dstrm}
\Jzg{0}{ab}\ds &=& \frac12\Ga^{ab}\ds,\nn\\
\Jzg{1}{a_1a_2a_3}\ds &=& \frac12\Ga^{a_1a_2a_3}\ds,\nn\\
\Jzg{2}{a_1\ldots a_6}\ds &=& \frac12\Ga^{a_1\ldots a_6}\ds,\nn\\
\Jzg{3}{a_0|a_1\ldots a_8}\ds &=& 4\de^{a_0[a_1}\Ga^{a_2\ldots a_8]}\ds,
\ee
where $\Ga^a$ are the ten real, symmetric $(32\times 32)$
$\Ga$-matrices of $SO(10)\subset GL(10)$ (see appendix \ref{appA}) and
$\Ga^{ab}=\Ga^{[a}\Ga^{b]}$ etc. denote their anti-symmetrised
products. Note that only the $SO(10)$ trace part of $\Jzg{3}{a_0|a_1\ldots
  a_8}$ is realised non-trivially, in accordance with the fact that
no Young tableaux other than fully antisymmetric ones can be built
with $\Ga$-matrices. Furthermore, we have rescaled the `level' 3 generator
by a factor $1/3$ relative to \cite{DaNi04,DaKlNi06a,DaKlNi06b}. As
emphasized in \cite{DaKlNi06a,dBHP06,DaKlNi06b}, the above
representation is {\em unfaithful} as the infinite-dimensional
group is realized on a finite number of spinor components.

Before proceeding it is useful to define the matrix
\be \label{gammastern}
\Ga^* := \Ga^1\Ga^0,
\ee
in terms of which the following relation holds for the
$(32\times 32)$ $\Ga$-matrices
\be\label{eps9}
\Ga^{\alpha_1\ldots \alpha_9}=\eps^{\alpha_1\ldots \alpha_9}\Ga^*
\quad\Rightarrow \quad \Ga^{\al_1\ldots  \al_k}=\frac{(-1)^{k(k-1)/2}}{(9-k)!}
\eps^{\al_1\ldots\al_k\beta_{k+1}\ldots\beta_9}\Ga_{\beta_{k+1}\ldots
  \beta_9}\Ga^*
\ee
with the $SO(9)$ invariant tensor $\eps^{\al_1\dots \al_9}$. The matrix $\Ga^*$
satisfies $(\Ga^*)^2=1$ and commutes with all $\Ga^\al$ for $\al=2,\ldots,10$,
but anticommutes with $\Ga^0$ and $\Ga^1$, and hence should be identified
with the chirality (helicity) matrix in $(1+1)$ space-time dimensions. By
defining $\chi_\pm=\frac12(1\pm\Ga^*)\chi$ for any 32-component spinor, it
therefore serves to split any such $\chi$ into two sets of 16-component
objects, which can be viewed as the right- and left-handed components,
respectively, of a spinor in (1+1) dimensions, and whose 16 `internal'
components transform as spinors under $SO(9)= K(SL(9))\subset\KEn$.

The (unfaithful) action of $\KEn$ on a Dirac-spinor $\ds$ is obtained
from (\ref{dstrm}) by restricting the range of the indices, as described
before. From the construction of the consistent representation we can in this
case derive a closed formula for the action of {\em all} $\KEn$ generators
by repeated commutation of the low level elements (\ref{dstrm}) and
use of (\ref{eps9}), and finally comparison with (\ref{ke9rels}).
The result is\footnote{The rescaling of the level $\ell=3$ generators
 by $1/3$ in comparison with \cite{DaNi04} is needed to ensure that
the level $(3k)$ generators are uniformly normalised, cf. also (\ref{normierung}).}
\be\label{ds9}
\Jg{3k}{\al\beta} &=& \frac12\Ga^{\al\beta}(\Ga^*)^k,\nn\\
\Jg{3k+1}{\al_1\al_2\al_3} &=& \frac12\Ga^{\al_1\al_2\al_3}(\Ga^*)^k,\nn\\
\Jg{3k+2}{\al_1\ldots \al_6} &=& \frac12\Ga^{\al_1\ldots
  \al_6}(\Ga^*)^k,\nn\\
\Jgb{3k+3}{\al\beta} &=& 0,
\ee
where, of course, $k\geq 0$. It follows from (\ref{ds9}) in particular
that $\Jgb{3k+3}{\al\beta}$ is represented trivially on the Dirac spinor,
and likewise that the relations involving $\Jgb{3k+3}{\al\beta}$ all
trivialise, as it should be. For the (reducible) Dirac representation,
we thus read off the relations (again for $k\geq 0$)
\be\label{dsideal}
\Jg{3k}{\al\beta} &=& \Jg{3k+6}{\al\beta}\; , \qquad
\Jgb{3k+3}{\al\beta} = 0 \;, \nn\\
\Jg{3k+1}{\al_1\al_2 \al_3} &=& -\frac1{6!}
    \eps^{\al_1\al_2 \al_3\beta_1\ldots \beta_6}
    \Jg{3k+5}{\beta_1\dots\beta_6},\nn\\
\Jg{3k+2}{\al_1\ldots \al_6} &=&
-\frac1{3!}\eps^{\al_1\ldots\al_6\beta_1\beta_2\beta_3}
\Jg{3k+4}{\beta_1\beta_2\beta_3}.
\ee
The existence of a  $32$-dimensional unfaithful representation of $\KEn$
(derived from the $32$-dimensional irreducible Dirac spinor of $\KEz$)
is thus reflected by the existence of a non-trivial ideal within the
Lie algebra $\KEn$, via (\ref{id}). For obvious reasons, we
will refer to this ideal as the {\em Dirac ideal} and designate it
by $\mf{i}_{\text{Dirac}}$. To be completely precise, the latter
is defined as the linear span within $\KEn$ of the relations (\ref{dsideal}).
It is straightforward to check that $\mf{i}_{\text{Dirac}}$ is indeed
an ideal, {\it i.e.} $[\KEn , \mf{i}_{\text{Dirac}}]\subset
\mf{i}_{\text{Dirac}}$. Furthermore, since by (\ref{dsideal}) all
generators of level greater than three can be expressed in terms of
lower level generators, the codimension of this ideal within
$\KEn$ is finite, and equal to the number of independent non-zero
elements up to level three, which is $2\times (36 + 84)$.
The resulting quotient is a finite-dimensional subalgebra of
$\mf{gl}(32)$ and has the structure
\be\label{quotient1}
\mf{q}_{\text{Dirac}} = \KEn/\mf{i}_{\text{Dirac}} =
  \mf{so}(16)_+ \oplus \mf{so}(16)_-.
\ee
To see that the Lie algebra on the r.h.s. has been correctly identified,
recall from \cite{DaKlNi06a,DaKlNi06b} that the quotient
algebra associated with the unfaithful Dirac-spinor
in $\KEz$ is $\mf{so}(32)$; according to (\ref{quotient1}) this splits into
$\mf{so}(16)_+\oplus\mf{so}(16)_-$, since all anti-symmetric $(16\times
16)$ matrices are contained in the list (\ref{ds9}).

Since $\Ga^*$ commutes with all these representation matrices, we can
decompose the $32$-dimensional $\KEn$ representation space further into
eigenspaces of $\Ga^*$ which are invariant under the $\KEn$ action.
These are projected out by $\frac12(1\pm\Ga^*)$, and we have
the branching
\be
{\bf 32} \quad\rightarrow\quad {\bf 16}_+ \oplus {\bf 16}_-
\ee
into two inequivalent
spinor representations of $\KEn$.
On the ${\bf  16}_\pm$ representations of $\KEn$, one can thus replace
$\Ga^*$ by $\pm 1$. This allows us to enlarge the Dirac ideal
(\ref{dsideal}) in two possible ways by replacing the relations
(\ref{dsideal}) by
\be\label{16pmrels}
\Jg{3k}{\al\beta} &=& \pm\Jg{3k+3}{\al\beta},\nn\\
\Jg{3k+1}{\al_1\al_2 \al_3} &=& \mp\frac1{6!}
    \eps^{\al_1\al_2 \al_3\beta_1\ldots \beta_6}
    \Jg{3k+2}{\beta_1\dots\beta_6},\nn\\
\Jgb{3k}{\al\beta} &=& 0,
\ee
for the ${\bf  16}_\pm$ representations, thereby defining two new
ideals $\mf{i}_{\text{Dirac}}^\pm \supset \mf{i}_{\text{Dirac}}$.
The quotient algebras are easily seen to be
\be\label{quotient2}
\mf{q}_{\text{Dirac}}^\pm = \KEn/\mf{i}_{\text{Dirac}}^\pm =
  \mf{so}(16)_\pm.
\ee

Let us now study in a bit more detail the ideal associated with the ${\bf
  16}_\pm$ Dirac-spinors of $\KEn$ determined by (\ref{16pmrels}) and,
in particular,  its orthogonal complement with respect to the $\KEn$
(and $E_9$ \cite{Ka90}) invariant form $\langle\cdot|\cdot\rangle$ under which
\be
\llangle\Jg{3k}{\al\beta}\Big|\Jg{3k}{\ga\de}\rrangle &=&
  -2\cdot 2!\,\de^{\al\beta}_{\ga\de} \qquad \left[ \;
  = \frac1{16}\tr\left(\Ga^{\al\beta}\Ga^{\ga\de}\right) \;\; \right],\nn\\
\llangle\Jg{3k+1}{\al_1\al_2\al_3}\Big| \Jg{3k+1}{\beta_1\beta_2\beta_3}
  \rrangle &=&
  -2\cdot3!\,\de^{\al_1\al_2\al_3}_{\beta_1\beta_2\beta_3} \qquad\left[ \;
  =\frac1{16}\tr\left(\Ga^{\al_1\al_2\al_3}
      \Ga^{\beta_1\beta_2\beta_3}\right) \;\right] ,\nn\\
\llangle\Jg{3k+2}{\al_1\ldots \al_6} \Big| \Jg{3k+2}{\beta_1\ldots \beta_6}
   \rrangle &=&
  -2\cdot 6!\,\de^{\al_1\ldots \al_6}_{\beta_1\ldots \beta_6} \qquad\left[ \;
  = \frac1{16}\tr\left(\Ga^{\al_1\ldots \al_6}\Ga^{\beta_1\ldots
    \beta_6}\right) \; \right]. \label{normierung}
\ee
We also have the consistency of orthogonality relations
\be
\llangle\Jg{3k+1}{\al_1\al_2\al_3}\Big|
  \Jg{3k+2}{\beta_1\ldots \beta_6}\rrangle=0 \qquad\left[ \; =
\frac1{16}\tr\left(\Ga^{\al_1\al_2\al_3}\Ga^{\beta_1\ldots \beta_6}\right)\;
\right].
\ee
Note that the invariant inner product $\tr$ on the $32$-dimensional
representation  agrees with the one on the algebra for the $\Jg{m}{}$
generators. Evaluated on $\Jgb{3k}{\al\beta}$ it vanishes in contrast
with the non-vanishing inner product in $\KEn$. This is no
contradiction since we are dealing with an unfaithful representation.

Defining the infinite linear combinations
\be\label{orth}
{\cal J}_\pm^{\al\beta} &=& \sum_{n\geq 0} (\pm 1)^n \Jg{3n}{\al\beta}, \nn\\
{\cal J}_\pm^{\al\beta\ga} &=&
   \sum_{n\geq 0} (\pm 1)^n \big( \Jg{3n +1}{\al\beta\ga} \pm
   \eps^{\al\beta\ga\delta_1 \dots \delta_6}
    \Jg{3n+2}{\delta_1\dots \delta_6} \big),
\ee
one checks that w.r.t. (\ref{Hilbert}),
\be
\left( {\cal J}_\pm^{\al\beta}\Big| Z \right) =
\left( {\cal J}^{\al\beta\ga}_\pm \Big| Z \right) = 0
\qquad \mbox{for all $Z\in \mf{i}_{\text{Dirac}}^\pm$},
\ee
and so ${\cal J}_\pm^{\al\beta}$ and ${\cal J}_\pm^{\al\beta\ga}$ formally
belong to the orthogonal complement of
   $\mf{i}_{\text{Dirac}}^\pm$.\footnote{Where the elements of
   $Z\in\mf{i}^\pm_{\text{Dirac}}$ are understood to be {\em finite} linear
   combinations of  (\ref{16pmrels}).} Thus, the elements
(\ref{orth}) are not proper elements of the vector space underlying the Lie
algebra $\KEn$ because the infinite series (\ref{orth}) do not converge in
the (Hilbert space) completion of $\KEn$ w.r.t. the norm (\ref{Hilbert}).
However, they do exist as {\em distributions}, that is, as elements of
the dual of the space of finite linear combinations of basis elements
(\ref{ds9}) (which is dense in the Hilbert space completion of $\KEn$).
This is also the reason why the elements $\{{\cal J}_\pm^{\al\beta},
{\cal J}_\pm^{\al\beta\ga}\}$ do not close into a proper subalgebra of
$\KEn$, as would be the case for the orthogonal complement of an ideal
in a finite-dimensional Lie algebra. Nevertheless, as we saw above,
there is a way to make sense of (\ref{orth})
as defining a Lie algebra by passing to the quotient algebras
(\ref{quotient1}) and (\ref{quotient2}). In section~\ref{currtrm}
we will see that these quotient algebras correspond to {\em generalised
evaluation maps} in terms of a loop algebra description.
The {\em distributional nature}
of these objects is also evident from the fact that formal commutation
of the elements (\ref{orth}) leads to infinite factors
$\sim \sum_{k=1}^\infty 1$. Whereas for $\KEn$ the
distributional nature can be made precise in terms of usual $\de$-functions
on the spectral parameter plane (see section~\ref{currtrm}), such a description
is not readily available for $\KEz$. Giving a more precise definition
of the space of distributions for $\KEz$ could prove helpful in understanding
the $\KEz$ structure better.

It may seem surprising that $\KEn$ admits non-trivial ideals, whereas $E_{9}$
does not (except for the one-dimensional center). One reason that $E_{9}$
does not admit any other ideals is the presence of the derivation $d$
as an element of $E_9$ (or any other affine) Lie algebra: because
relations such as (\ref{dsideal}) and (\ref{16pmrels}) involve {\em
  different} affine levels (even within generators $J_{(n)}$ of fixed
$\mf{sl}(9)$ level $n$, as we saw), commutation with $d$ will change the relative
coefficients between the terms defining the ideal by (\ref{E9aff}),
hence will force the individual terms to vanish also, thus leading to
the trivial ideal $\mf{i}=0$. The existence of non-trivial ideals in
$\KEn$ is thus due in particular to the fact that $d$ is {\em not} an
element of $\KEn$. In the section~\ref{currtrm} we shall give a loop
algebra interpretation of this result.

\end{subsection}

\begin{subsection}{Vector-spinor ideal}
\label{vssec}

The $\KEz$ transformation of the $320$-component vector-spinor
$\psicomp_a$ can also be written in terms of $SO(10)$ $\Ga$-matrices
\cite{DaKlNi06a,dBHP06}. For the first three $SO(10)$ `levels' the
$\KEz$ expressions are\footnote{When comparing these expressions
  to \cite{DaKlNi06a} we recall once more that we have re-scaled $\Jzg{3}{}$
  by $1/3$ as for the Dirac-spinor.}
\be
(\Jzg{0}{ab}\psicomp)_c &=& \frac12\Ga^{ab}\psicomp_c
  +2\de^{[a}_c\psicomp^{b]},\nn\\
(\Jzg{1}{a_1a_2a_3}\psicomp)_b &=& \frac12\Ga^{a_1a_2a_3}\psicomp_b
  +4\de_b^{[a_1}\Ga^{a_2}\psicomp^{a_3]}
  -\Ga_b{}^{[a_1a_2}\psicomp^{a_3]},\nn\\
(\Jzg{2}{a_1\ldots a_6} \psicomp)_b &=& \frac12\Ga^{a_1\ldots
  a_6}\psicomp_b
  -10 \de_b^{[a_1}\Ga^{a_2\ldots a_5}\psicomp^{a_6]}
  +4 \Ga_b{}^{[a_1\ldots a_5}\psicomp^{a_6]},\nn\\
(\Jzg{3}{a_0|a_1\ldots a_8} \psicomp)_b &=&
  \frac{16}9\left(\Ga_b{}^{a_1\ldots a_8}\psicomp^{a_0} -
  \Ga_b{}^{a_0[a_1\ldots a_7}\psicomp^{a_8]}\right)\nn\\
&& +4\de^{a_0[a_1}\Ga^{a_2\ldots a_8]}\psicomp_b
  -56 \de^{a_0[a_1}\Ga_b{}^{a_2\ldots a_7}\psicomp^{a_8]}\\
&& +\frac{16}9\left(8\de_b^{a_0}\Ga^{[a_1\ldots a_7}\psicomp^{a_8]}
   - \de_b^{[a_1}\Ga^{a_2\ldots a_8]}\psicomp^{a_0}
   + 7 \de_b^{[a_1}\Ga_{a_0}{}^{a_2\ldots a_7}\psicomp^{a_8]}\right).\nn
\ee
Reducing these transformations to $\KEn$ one decomposes the gravitino field $\psicomp_a$
into an $SO(9)$ vector spinor $\psicomp_\al$, and in addition the component $\psicomp_{1}$
entering via
\be
\psione:=\Ga^1\psicomp_1,
\ee
which transforms in
the spinor representation of the two-dimensional Lorentz group $SO(1,1)$ and $SO(9)\subset\KEn$. The correspondence of the fields
$\psicomp_\al$ and $\psione$ with the fermionic fields used in \cite{NiSa05}
will be explained in section~\ref{currtrm}.

Computing the $\KEn$ transformations for `levels' $0$ up to $3$
on the components $\psicomp_\al$ one obtains
\be\label{ke9trmal}
(\Jg{0}{\al\beta}\psicomp)_\ga &=&
  \frac12\Ga^{\al\beta}\psicomp_\ga
    + 2\de^{[\al}_{\ga}\psicomp^{\beta]},\nn\\
(\Jg{1}{\al_1\al_2\al_3}\psicomp)_\beta &=&
  \frac12\Ga^{\al_1\al_2\al_3}
    \psicomp_\beta
  + 4 \de_\beta^{[\al_1} \Ga^{\al_2} \psicomp^{\al_3]}
  - \Ga_\beta{}^{[\al_1\al_2}    \psicomp^{\al_3]} ,\nn\\
(\Jg{2}{\al_1\ldots \al_6} \psicomp)_\beta &=&
  \frac12\Ga^{\al_1\ldots  \al_6}\psicomp_\beta
  -10 \de_\beta^{[\al_1}\Ga^{\al_2\ldots \al_5}\psicomp^{\al_6]}
  +4 \Ga_\beta{}^{[\al_1\ldots \al_5}\psicomp^{\al_6]},\nn\\
(\Jg{3}{\al\beta}\psicomp)_\ga &=& -\Ga^* \left[
  \frac12\Ga^{\al\beta}\psicomp_\ga
    + 2\de^{[\al}_{\ga}\psicomp^{\beta]}\right],\nn\\
(\Jgb{3}{\al\beta}\psicomp)_\ga &=& -\Ga^* \left[
  \frac29\de^{\al\beta}\Ga_\ga-2\de^{(\al}_\ga\Ga^{\beta)}\right]\Ga^\de\psicomp_\de.
\ee
Note that the transformations on
$\psicomp_\al$ close on themselves.
Extending the action (\ref{ke9trmal}) by the commutation
relations (\ref{ke9rels}) we deduce the general action on 
$\psicomp_\al$, 
\be\label{ke9trmalfull}
(\Jg{3k}{\al\beta}\psicomp)_\ga &=& \left(-\Ga^*\right)^k
  \left[\frac12\Ga^{\al\beta}\psicomp_\ga
   +2\de^{[\al}_\ga\psicomp^{\beta]}\right],\nn\\
(\Jg{3k+1}{\al_1\al_2\al_3}\psicomp)_\beta &=& \left(-\Ga^*\right)^k
  \bigg[\frac12\Ga^{\al_1\al_2\al_3} \psicomp_\beta
  + 4 \de_\beta^{[\al_1} \Ga^{\al_2} \psicomp^{\al_3]}
  - \Ga_\beta{}^{[\al_1\al_2}    \psicomp^{\al_3]}\nn\\
&&\quad\quad\ \ +k\left(\frac13\Ga^{\al_1\al_2\al_3\beta}+2\de^{\beta [\al_1}\Ga^{\al_2\al_3]}\right)\Ga^\ga\psicomp_\ga
\bigg], \nn\\
(\Jg{3k+2}{\al_1\ldots \al_6}\psicomp)_\beta &=&
  \left(-\Ga^*\right)^k  \bigg[
  \frac12\Ga^{\al_1\ldots  \al_6}\psicomp_\beta
  -10 \de_\beta^{[\al_1}\Ga^{\al_2\ldots \al_5}\psicomp^{\al_6]}
  +4 \Ga_\beta{}^{[\al_1\ldots \al_5}\psicomp^{\al_6]}\nn\\
&&\quad\quad\ \ +k\left(\frac23\Ga^{\al_1\ldots \al_6\beta}-2\de^{\beta [\al_1}\Ga^{\al_2\dots\al_6]}\right)\Ga^\ga\psicomp_\ga
\bigg],\nn\\
(\Jgb{3k}{\al\beta}\psicomp)_\ga &=& \left(-\Ga^*\right)^k k \left[
  \frac29\de^{\al\beta}\Ga_\ga-2\de^{(\al}_\ga\Ga^{\beta)}\right]\Ga^\de\psicomp_\de.
\ee
Similar to (\ref{dsideal}) we immediately find the following relations
which are valid on the $\psicomp_\al$ components, 
\be\label{vsideal}
\Jg{3k}{\al\beta} &=& \Jg{3k+6}{\al\beta},\nn\\
\Jg{3k+7}{\al_1\al_2\al_3} - \Jg{3k+1}{\al_1\al_2\al_3}
 &=&\frac1{6!}\eps^{\al_1\al_2\al_3\beta_1\ldots\beta_6}\left(
     \Jg{3k+5}{\beta_1\ldots \beta_6}
    -  \Jg{3k-1}{\beta_1\ldots \beta_6}\right),\nn\\
(3k+1)\Jg{3k+7}{\al_1\al_2\al_3} - (3k+7) \Jg{3k+1}{\al_1\al_2\al_3}
 &=& -\frac1{6!}\eps^{\al_1\al_2\al_3\beta_1\ldots\beta_6}\nn\\
&& \quad\times\left(
   (3k-1)\Jg{3k+5}{\beta_1\ldots \beta_6}
   -(3k+5)\Jg{3k-1}{\beta_1\ldots \beta_6}\right),\nn\\
(k+2)\Jgb{3k}{\al\beta} &=& k\, \Jgb{3k+6}{\al\beta}.
\ee
The first two relations arise from considering the
$\psicomp_\al$ pieces of the transformed spinor (\ref{ke9trmalfull}),
the latter two can be derived by focussing on the trace parts
in the transformed spinor (\ref{ke9trmalfull}) and are evidently
$k$-dependent. Note also that the relations in the middle involve
{\em four} different $\mf{sl}(9)$ levels.

Just as in the Dirac case it follows immediately from the form of
the  transformations (\ref{ke9trmalfull}) 
that $\Ga^*$ commutes with all representation matrices and therefore
one can restrict to the $\Ga^*=\pm {\bf 1}$ eigenspaces. Hence, on
the $\Ga^*=\pm {\bf 1}$ eigenspaces the relations
(\ref{vsideal}) simplify in analogy with (\ref{16pmrels}) to
\be\label{vsidealpm}
\Jg{3k}{\al\beta} &=& \mp\Jg{3k+3}{\al\beta},\nn\\
\Jg{3k+4}{\al_1\al_2\al_3} \pm \Jg{3k+1}{\al_1\al_2\al_3}
 &=& \mp\frac1{6!}\eps^{\al_1\al_2\al_3\beta_1\ldots\beta_6}\left(
    \Jg{3k+5}{\beta_1\ldots \beta_6}
     \pm \Jg{3k+2}{\beta_1\ldots \beta_6}\right),\nn\\
(3k+1)\Jg{3k+4}{\al_1\al_2\al_3} \pm (3k+4)\Jg{3k+1}{\al_1\al_2\al_3}
  &=&\pm\frac1{6!}\eps^{\al_1\al_2\al_3\beta_1\ldots\beta_6}\nn\\
&&\quad\times\left( (3k+2)\Jg{3k+5}{\beta_1\ldots \beta_6}
     \pm (3k+5)\Jg{3k+2}{\beta_1\ldots \beta_6}\right),\nn\\
\Jgb{3k}{\al\beta} &=& (\pm 1)^{k+1}\,k \Jgb{3}{\al\beta}.
\ee
We stress that these and (\ref{vsideal}) are valid only on the
$\psicomp_\al$ components.

The transformation properties of the remaining component
$\psione=\Ga^1\psicomp_1$
are more complicated. At the first
three levels, they read
\be\label{ke9trm1}
\Jg{0}{\al\beta}\psione &=&
  \frac12\Ga^{\al\beta}\psione,\nn\\
\Jg{1}{\al_1\al_2\al_3}\psione &=&
  -\frac1{2} \Ga^{\al_1\al_2\al_3} \psione
  - \Ga^{[\al_1\al_2} \psicomp^{\al_3]},\nn\\
\Jg{2}{\al_1\ldots \al_6}\psione &=&
  \frac12\Ga^{\al_1\ldots\al_6}\psione
   +4\Ga^{[\al_1\ldots \al_5}\psicomp^{\al_6]},\nn\\
\Jg{3}{\al\beta}\psione &=&
   -\frac12\Ga^*\Ga^{\al\beta}\psione,\nn\\
\Jgb{3}{\al\beta}\psione &=&
  2 \Ga^*\Ga^{(\al}\psicomp^{\beta)},
\ee
where the mixing of $\psicomp_\al$  into $\psione$ 
is manifest.
We can again use the $\KEn$ commutation relations (\ref{ke9rels}) to
deduce the action for all generators from (\ref{ke9trm1}),
\be\label{ke9trm1full}
\Jg{3k}{\al\beta}\psione &=& (-\Ga^*)^k\bigg[
  \frac12\Ga^{\al\beta}\psione+k^2\Ga^{\al\beta}\Ga^\ga\psicomp_\ga\bigg]\nn\\
\Jg{3k+1}{\al_1\al_2\al_3}\psione &=& (-\Ga^*)^k \bigg[
  -\frac1{2} \Ga^{\al_1\al_2\al_3} \psione
  - (3k+1)\Ga^{[\al_1\al_2} \psicomp^{\al_3]}\nn\\
&&\quad\quad\quad\ \  -\frac13 k(3k+1)\Ga^{\al_1\al_2\al_3}\Ga^\beta\psicomp_\beta\bigg],\nn\\
\Jg{3k+2}{\al_1\ldots \al_6}\psione &=& (-\Ga^*)^k \left[
  \frac12\Ga^{\al_1\ldots\al_6}\psione
   +2(3k+2)\Ga^{[\al_1\ldots \al_5}\psicomp^{\al_6]}\right]\nn\\
&&\quad\quad\quad\ \  +\frac13 k(3k+2)\Ga^{\al_1\ldots \al_6}\Ga^\beta\psicomp_\beta\bigg],\nn\\
\Jgb{3k}{\al\beta}\psione &=&   (-\Ga^*)^k
  k \bigg[ -2\Ga^{(\al}\psicomp^{\beta)}+\frac29\de^{\al\beta}\Ga^\ga\psicomp_\ga\bigg].
\ee
Using (\ref{ke9trmalfull}) and (\ref{ke9trm1full}) we can now
deduce relations analogous to (\ref{vsideal})
valid on both the $\psicomp_\al$ and the
$\psione$ components of $\psi_a$ and hence on the full
representation. These define the vector-spinor ideal. Since the
$k$-dependence in (\ref{ke9trm1full}) is \textit{quadratic},
they will be more complicated than (\ref{vsideal}) and involve up to
\textit{six} different $\mf{sl}(9)$ levels. We will discuss their
structure at the end of this section and give them explicitly in a
simplifying `gauge' which we now present.

{}From the transformations (\ref{ke9trmalfull}) 
it can be shown
that the gamma-trace $\Ga^\al \psicomp_\al$ transforms into itself.
For this reason, we can consistently consider the tracelessness
condition
\be\label{trace}
\Ga^\al \psicomp_\al = 0,
\ee
which, as we will recall in section~\ref{currtrm}, corresponds to a
supersymmetric gauge choice for the dilatino in the reduction from
three to two dimensions.
As shown in \cite{KlNi06} and \cite{DaKlNi06b}, cf. eq.~(2.29),
this condition is compatible with $K(E_n)$ only for $n=9$, as
required. In fact it follows from (\ref{ke9trmalfull})
that $\Ga^\al \psicomp_\al$ transforms just as
a Dirac-spinor.
With the tracelessness condition (\ref{trace}), the $k$-dependence
in (\ref{ke9trmalfull}) vanishes,
and in particular $\Jgb{3k}{\al\beta}$ acts trivially on
$\psicomp_{\al}$ for all $k$.
The corresponding ideal would then be the same as in (\ref{dsideal}).
That is, we have the same
$SO(16)_+\times SO(16)_-$ acting on this part of the gravitino.
The action for $\Jg{3k+2}{\al_1\ldots\al_6}$ can be written in a dual
form as shown above. Moreover, we see that we can again specialise to
the $\Ga^*=\pm 1$ subspaces. There it is easiest to deduce the
following relations for the vector-spinor components $\psicomp_{\al}$ in the
traceless gauge,
\be\label{rsideal1}
\Jg{3k+3}{\al\beta} &=& \mp\Jg{3k}{\al\beta},\nn\\
\Jg{3k+1}{\al_1\al_2 \al_3} &=& \pm\frac1{6!}
    \eps^{\al_1\al_2 \al_3\beta_1\ldots \beta_6}
    \Jg{3k+2}{\beta_1\dots\beta_6},\nn\\
\Jgb{3k}{\al\beta} &=& 0
\ee
in analogy with (\ref{16pmrels}) (except that $\Ga^*$ is replaced by
$(-\Ga^*)$). By the
arguments of the preceding sections the relevant ideal on the
components $\psicomp_\al$ gives a quotient isomorphic to
$\mf{so}(16)_\pm$. However, as noted above, the component
$\psione$ {\em mixes} with the $\psicomp_\al$ components and one can
show that they cannot be decoupled by a change of basis. Therefore
the relations (\ref{rsideal1}) have to be weakened in order to
describe the full vector-spinor ideal. In the gauge (\ref{trace})
the transformations (\ref{ke9trm1full}) simplify and the
$k$-dependence becomes linear instead of quadratic. Then it is easy to check that
the vector-spinor ideal relations are
identical to (\ref{vsideal}).

Let us now summarize our findings and write out the branching of the
${\bf 320}$ representation of $\KEz$ into representations of its $\KEn$
subalgebra. In comparison with the Dirac representation, the
vector-spinor representation exhibits a curious new feature in the
branching. Namely, the transformations on $\psione$ contain
contributions also involving $\psicomp_\al$.
On the other hand the $\psicomp_\al$ components
transform solely among themselves. This means that the $\psicomp_a$
representation of $\KEz$ {\em does not completely reduce into irreducible
representations of $\KEn$} as one might have expected, rather we
have a triangular structure
\be\label{branching}
{\bf 320} \rightarrow  \big(  {\bf 16}_+
 \oplus {\bf 16}_- \big)+\big({\bf 128}_+ \oplus {\bf 128}_-\big) +
\big(  {\bf 16}_+
 \oplus {\bf 16}_- \big)
\ee
where the plus signs between the parantheses denote a \textit{semidirect} sum of, from right to left, the trace components $\Ga^\al\psicomp_\al$, the traceless part of $\psicomp_\al$, and the $\eta$ components: only the trace components transform among themselves, the other two summands mix with those to the left.
These results are
in accordance with the results of \cite{NiSa05}, see eqns.~(5.12)
there, as we will discuss in more detail below.
The triangular
structure can, for each chirality, be pictured by $\KEn$ representation matrices of
block form
\be
\left(\begin{array}{ccc}*&*&*\\0&*&*\\0&0&*\end{array}\right)\,.
\ee
The blocks are of dimensions $16\times 16$, $128\times 128$ and
$16\times 16$, respectively, and correspond to the summands in the
decomposition (\ref{branching}) in reverse order. 
In this manner,
the lower right block corresponds to the transformation of the
gamma-trace $\Ga^\al\psicomp_\al$ into itself. The
non-reducibility of the ${\bf 320}$ is tantamount to saying that
the representation matrix cannot be block-diagonalised.

The structure of the ideal in $\KEn$ associated with this
representation can be revealed by starting with the `innermost'
layer of the triangular structure, namely $\Ga^\al\psicomp_\al$.
As stated above this transforms as a Dirac-spinor so the
associated quotient algebra (projected onto the two $\Ga^*$ chiralities) is
$\mf{so}(16)_\pm$, cf. (\ref{quotient2}). This gets enlarged since the ideal relations are
weakened due to the appearance of the gamma-trace in the $\KEn$
action on $\psicomp_\al$, cf. (\ref{ke9trmalfull}), and even more
due to (\ref{ke9trm1full}). The expected structure is
\be
\mf{q}^\pm_{\text{vs}} = \mf{so}(16)_\pm + \mf{p}^{(1)}_\pm +
\mf{p}^{(2)}_\pm
\subset \mf{gl}(160)
\ee
as a semi-direct sum with actions from left to right as before, so
that $\mf{so}(16)_\pm$ acts on the pieces $\mf{p}^{(1)}_\pm$ and
$\mf{p}^{(2)}_\pm$ via some representation,
$[\mf{p}^{(1)}_\pm,\mf{p}^{(1)}_\pm]\subset \mf{p}^{(2)}_\pm$, and
$\mf{p}^{(2)}_\pm$ is abelian. In the tracelass case (\ref{trace})
this can be evaluated further and we find
\be
\mf{q}^\pm_{\text{vs}} = \mf{so}(16)_\pm +\mf{p}^{}_\pm
\subset \mf{gl}(144) \quad\quad(\Ga^\al\psicomp_\al=0)
\ee
where $\mf{p}^{}_\pm$ are $128$ abelian translations and the
whole ideal has codimension $248$ as can be counted from
(\ref{vsideal}): The action of all $\KEn$ generators in the
vector-spinor representation can be reduced to that of
$\Jg{0}{\al\beta}$, $\Jg{1}{\al_1\al_2\al_3}$, $\Jg{2}{\al_1\ldots
  \al_6}$ and $\Jgb{3}{\al\beta}$ which amount to
$(40+80)+(80+48)=120+128$ independent generators. Via the relations
(\ref{vsideal}), all higher
level generators can thus be expressed as linear combinations of these
$248$ basic ones. This discussion shows that the structure of the
ideals in the vector-spinor case is far richer than that of the
Dirac-spinor.

\end{subsection}

\end{section}

\begin{section}{Relation to current algebra realisation}
\label{ke9curr}

In previous work \cite{NiSa05}, $\KEn$ transformations of unfaithful
fermion representations were derived starting from the linear system
description of $N=16$ supergravity in $D=2$ \cite{Ni87,NiSa98}. In the present
section we will show that the transformations (\ref{ke9trmalfull}) and
(\ref{ke9trm1full}) we deduced from the dimensionally reduced theory
above are completely equivalent to those in the linear system.

\begin{subsection}{$\mf{so}(16)\subset E_{8(8)}$}
\label{soapp}

Since the linear system transformations are written using the spectral
parameter presentation of $\KEn$ in the $K(E_8)\equiv\mf{so}(16)$
decomposition of $E_8$ we first need to briefly recall some notation
necessary for the comparison; in particular, we require the $E_8$
commutation relations adapted to the compact $\mf{so}(16)$ subalgebra.
In this basis, $E_{8(8)}$ decomposes into the adjoint ${\bf 120}$ of
$\mf{so}(16)$ (corresponding to the anti-symmetric compact generators)
and the $\mf{so}(16)$ spinor representation $\bf{128_s}$ (corresponding
to the symmetric non-compact generators) which can be further
decomposed as
\begin{align}\label{A1}
X^{IJ} \in {\bf 120}\;&\rightarrow&{\bf
(28,\overline{1}) \oplus (1,\overline{28})\oplus (8_s,\overline{8}_c)}
   \rightarrow&\quad {\bf 28 \oplus 28 \oplus 56_v \oplus 8_v},\\
Y^A \in {\bf 128_s} &\rightarrow& {\bf (8_v,\overline{8}_v) \oplus (8_s,\overline{8}_c)}
  \rightarrow&\quad {\bf 1}\oplus{\bf 28}\oplus{\bf 35_v}\oplus{\bf 8_v}\oplus{\bf 56_v},\nn
\end{align}
with the chain of embeddings $\mf{so}(16)\supset\mf{so}(8)\oplus\mf{so}(8)
\supset\mf{so}(8)$, where the indices $v,s,c$ (= vector, spinor, and
conjugate spinor) label the three inequivalent eight-dimensional
representations of the various $SO(8)$ groups. The diagonal subalgebra
$\mf{so}(8)$ is to be identified with the $\mf{so}(8)\subset\mf{sl}(8)$
of the preceding sections. Furthermore, we here take over the notation
from \cite{NiSa05}: $I,J=1,\ldots,16$ are $SO(16)$ vector indices and
$A=1,\ldots,128$ labels the components of a chiral $SO(16)$ spinor.
Evidently, the first line in (\ref{A1}) corresponds to the $\mf{so}(8)$
representations inherited from table~\ref{tabe8a7}. The formulas relating the $SO(9)$ and $SO(16)$ bases
are spelled out in
appendix \ref{appB}.
From (\ref{A1})
we also recover the decompositions of $SO(16)$ under its $SO(9)$
subgroup, viz.
\be
\bf{120} \; \rightarrow \; \bf{36} \oplus \bf{84} \;\; , \quad
\bf{128_s} \; \rightarrow \; \bf{44} \oplus \bf{84}.
\ee
In the conventions of \cite{KoNiSa00}, the $E_8$ commutation relations read
\be \label{so16kr}
[ X^{IJ} , X^{KL} ] &=& 2\delta^{I[K} X^{L]J} - 2\delta^{J[K} X^{L]I}, \nn\\
{}[ X^{IJ} , Y^A ]    &=& -\frac12 \Ga^{IJ}_{AB} Y^B \quad , \qquad
[Y^A , Y^B ] = \frac14 \Ga^{IJ}_{AB} X^{IJ}.
\ee
With the current algebra generators (for $m\in\Z$)
\be
\Xso{m}{IJ} \equiv X^{IJ} \otimes t^m \quad , \qquad
\Yso{m}{A} \equiv Y^A \otimes t^m,
\ee
the $\KEn$ generators can be represented in the form (for $m\geq 0$)
\be
\Aso{m}{IJ} := \frac12 \left( \Xso{m}{IJ} + \Xso{-m}{IJ}\right) \quad , \qquad
\Sso{m}{A} := \frac12 \left( \Yso{m}{A} - \Yso{-m}{A}\right),
\ee
implying $\Sso{0}{A}\equiv 0$. The $\KEn$ commutation relations then read
\be\label{ke9so16}
[ \Aso{m}{IJ} , \Aso{n}{KL} ] &=& 2 \delta^{[I[K} \Big( \Aso{m+n}{L]J]}
       + \Aso{|m-n|}{L]J]} \Big) \nn\\
{}[ \Aso{m}{IJ} , \Sso{n}{A} ]    &=& -\frac14 \Ga^{IJ}_{AB}
         \Big( \Sso{m+n}{B} - {\rm sgn} (m-n) \Sso{|m-n|}{B} \Big) \nn\\
{} [\Sso{m}{A} , \Sso{n}{B} ] &=& \frac18 \Ga^{IJ}_{AB} \Big( \Aso{m+n}{IJ}
    - \Aso{|m-n|}{IJ} \Big)
\ee
for $m,n \geq 0$ (recall that the central term drops out).

In the formulation (\ref{ke9so16}) we can immediately look for ideals
of $\KEn$.
The Dirac ideals $\mf{i}_{\text{Dirac}}^\pm$ are now defined by the
relations
\be\label{dsidcurr}
\Aso{m}{IJ} - (\pm 1)^m \Aso{0}{IJ} = 0 \quad , \qquad \Sso{m}{A} =0.
\ee
That is, the ideals are defined as the linear span of the expressions
on the l.h.s., and it is then straightforward to verify the ideal property,
namely that these subspaces are mapped onto themselves under the adjoint
action of $\KEn$. The quotient algebras obtained by division of $\KEn$ by
these ideals are obviously isomorphic to $\mf{so}(16)$ for both
choices of signs.

The vector-spinor ideals $\mf{i}_{\text{vs}}^\pm$, on the other hand,
can be defined by the relations (for $m\geq 1$)
\be
\Aso{m}{IJ} - (\pm 1)^m \Aso{0}{IJ} = 0 \quad , \qquad
\Sso{m}{A} \mp (\pm 1)^{m} m \Sso{1}{A} = 0.
\ee
They define smaller ideals of codimension $248$ since everything is
determined by $\Aso{0}{IJ}$ and $\Sso{1}{A}$. The part of the
above relations involving $\Aso{m}{IJ}$ is identical to that of
the Dirac-spinor (\ref{dsidcurr}) indicating that there is some
relation of the associated quotient to $\mf{so}(16)$ with an additional part
arising from the $\Sso{m}{A}$ relations. We will see this in more
detail below.

The vector-spinor ideals $\mf{i}_{\text{vs}}^\pm$ can be generated
from $\Aso{1}{IJ} \mp \Aso{0}{IJ}=0$ since for example
\be
\left[\Aso{1}{IJ} \mp \Aso{0}{IJ}, \Sso{1}{A}\right] =
  -\frac14\Ga^{IJ}_{AB}\left(\Sso{2}{B}\mp 2\Sso{1}{B}\right)
\ee
implies by the ideal property that $\Sso{2}{B}\mp 2\Sso{1}{B}$ has to
vanish. Similar calculations show that $\Aso{1}{IJ} \mp \Aso{0}{IJ}=0$
generates all ideal relations.

In this basis it is not hard to construct further ideals.
One example is obtained by starting from the
relation $\Sso{2}{A} \mp 2\Sso{1}{A}=0$, without requiring that
$\Aso{1}{IJ} \mp \Aso{0}{IJ}=0$. Commuting with
$\Sso{1}{B}$ and demanding that the resulting expression also belongs
to the ideal leads to
\be
\Aso{3}{IJ}-\Aso{1}{IJ}\mp2 \Aso{1}{IJ} \pm
\Aso{0}{IJ} = 0,
\ee
a relation involving four affine levels. In the case of the
vector-spinor these vanish by taking pairwise combinations, here they
define a new ideal which is strictly smaller than the vector-spinor
ideal.

In section~\ref{dssec} we explained that the absence of non-trivial ideals in $E_9$
(other than the one-dimensional center)
can be interpreted as a consequence of the presence of the derivation
$d$. In the current algebra realization, $d$ acts by differentiation:
$d\equiv \partial_t$. Setting $X(t_0) = 0$ for some fixed $t_0$ would
then force all higher repeated commutators of this element with $d$ to vanish
at  $t=t_0$ by consistency. This, in turn, would imply the vanishing of
all derivatives $\partial_t^n X(t_0)$, hence would force $X(t)=0$ (assuming
analyticity in $t$). This confirms again that the existence of non-trivial
ideals  in $\KEn$ is thus due in particular to the fact that $d$ is
not an  element of $\KEn$. The orthogonal complement of the ideal, given
formally by (\ref{orth}), corresponds to distributions
$X(t) = X_0 \,\delta(t - t_0)$ where, as we will see presently, $t_0=\pm 1$.
The associated ideal then consists of all elements of the loop algebra which
vanish at those points. We stress that this requires studying a distribution
space outside of $\KEn$ and that this could prove a useful strategy
also for further investigations of $\KEz$.

\end{subsection}

\begin{subsection}{Current algebra fermion transformations}
\label{currtrm}

In \cite{NiSa05} it was realised that in the linear systems approach
to two-dimensional $N=16$ supergravity the transformation rules for
the fermions 
can be written succinctly in terms of a
current algebra description with a current parameter $t$.
The non-propagating fermions are the gravitino $\varphi^I$ and the
dilatino $\varphi_2^I$, coming from the gravitino
in three dimensions.\footnote{These are called $\psicomp^I$
 and $\psicomp^I_2$ in \cite{NiSa05}, but we choose a different notation here to
 avoid confusion with the gravitino in 11-dimensional supergravity.}
They both transform in the vector representation of $SO(16)$,
while the field $\chi^{\dot{A}}$ accomodates the $128$ physical
fermions and transforms in the
conjugate spinor representation of $SO(16)$.
The dilatino $\varphi_2^I$ can be
gauged away by use of local supersymmetry \cite{NiSa05}, corresponding to
the tracelessness condition (\ref{trace}). It follows from a
comparison with the reduction of 11-dimensional supergravity to
three dimensions \cite{Nicolai:1986jk}
that the correspondence between these $SO(16)$ representations and
those used in the foregoing sections is (modulo a factor 2 in the relative normalisation of $\chi^{\dot{A}}$ compared to $\psicompt_2^I$ and $\psicompt^I$,
required for the canonical normalisation of the Dirac term)
\be\label{lsferm}
\chi^{\dot{A}} &\leftrightarrow& \psicomp_i - \frac12 \Ga_i\Ga^j\psicomp_j, \nn\\
\psicompt_2^I &\leftrightarrow&  \Ga^* (\Ga^2\psicomp_2 +
\Ga^i\psicomp_i),\nn\\
\psicompt^I &\leftrightarrow& -\Ga^1\psicomp_1-\Ga^i\psicomp_i,
\ee
thus breaking $SO(9)$ covariance down to $SO(8)$. 
We have suppressed the two-dimensional Dirac-spinor indices on the l.h.s
(which take two values, so that e.g. the $\psicompt^I$ stands for
$2\times 16$ components $\psicompt^I_{\pm}$), and 
the $SO(9)$ spinor indices on the r.h.s (of which there are
$2\times 16$, giving $2 \times 128$ components for the first line).
Thus the number of components on both sides match.

The most general $\KEn$ Lie algebra element can be written in
the form \cite{NiSa05}\footnote{Since we are interested for the moment
  in the purely algebraic aspects of the transformation we suppress
  the space-time dependence throughout. (The spectral parameter $t$
  also depends on two-dimensional space-time.)}
\be
\Ups(t) &=& 
\frac12\sum_{n=0}^\infty
  \Ups_n^{IJ} X^{IJ}\otimes (t^{-n}+t^n)
    + \sum_{n=1}^\infty \Ups_n^A Y^A \otimes (t^{-n}-t^n) \nn\\
&\equiv& \frac12 \Ups^{IJ}(t) X^{IJ} + \Ups^A(t) Y^A.
\ee
It can then be shown that $\KEn$ acts on the chiral components of
the fermions via evaluation at the points $t =\pm 1$ in the spectral
parameter plane (cf.
eqn.~(5.12) of \cite{NiSa05})\footnote{We note that in \cite{NiSa05}
  it was also shown that, considering only induced $\KEn$ transformations,
  there is a non-linear combination of the fermionic and bosonic fields
  that reduces this action to an action of $SO(16)_+\times SO(16)_-$.} as
\be\label{lstrm}
\de_\Ups \psicompt_{2\pm}^I &=& \psicompt_{2\pm}^J \Ups^{IJ}|_{t=\mp
  1},\nn\\
\de_\Ups \chi^{\dot{A}}_\pm &=&
  \frac14\Ga^{IJ}_{\dot{A}\dot{B}}\chi^{\dot{B}}_\pm\Ups^{IJ}|_{t=\mp 1}
     -\Ga^I_{A\dot{A}}\psicompt_{2\pm}^I\del_t\Ups^A|_{t=\mp 1},\\
\de_\Ups \psicompt_\pm^I &=& \psicompt_\pm^J\Ups^{IJ}|_{t=\mp 1}
   \pm\Ga_{A\dot{B}}^I\chi_\pm^{\dot{B}}\del_t\Ups^A|_{t=\mp 1}
  \mp 2 \psicompt_{2\pm}^J
  \del_t^2\Ups^{IJ}
|_{t=\mp 1}.\nn
\ee
Thus, from the point of view \cite{NiSa05}
the action of $\KEn$ on the fermions can be viewed as an
{\em evaluation map} of the $\KEn$ elements, not at the origin
in spectral parameter space $t=0$ but at $t=\pm 1$. In fact, we
are dealing with a {\em generalised} evaluation map in that the
transformations depend on up to second derivatives in the spectral
parameter at the points $t=\pm 1$.

Now we compare (\ref{lstrm}) to (\ref{ke9trmalfull}) and (\ref{ke9trm1full}).
Writing
\be\label{hat1}
\Ups^{IJ}(t)|_{t=\pm 1} &=&
2 \sum_{n=0}^\infty \Ups_n^{IJ}(\pm 1)^n
\ee
suggests the structure
of an $\mf{so}(16)$, so we see that the Taylor expansion (\ref{hat1}) (considered as a formal power series) 
should indeed be identified
with the formal infinite sum in (\ref{orth}). Considering also the parameters $\partial_t \Ups^A$ and $\partial^2_t \Ups^{IJ}$, we can see that there is a
structural agreement between the transformations (\ref{lstrm})
and those of the vector-spinor ($\psicomp_\al,\psione$) in
section~\ref{vssec}. To make the agreement exact,
we rewrite (\ref{lstrm}) using the basis given in the preceding section, and $\Ga^*$ as the chirality (helicity) matrix in (1+1) spacetime dimensions,
\be \label{omskrivet1}
(A^{(m)KL}{\psicompt_2})^I &=& 2(-\Ga^{\ast})^m\delta^{I[K}{\psicompt_2}^{L]},\nonumber\\
(A^{(m)KL}{\psicompt})^I &=& 2(-\Ga^{\ast})^m\delta^{I[K}{\psicompt}^{L]}+4m^2(-\Ga^{\ast})^{m-1}\delta^{I[K}{\psicompt_2}^{L]},\nonumber\\
(A^{(m)KL}{\chi})^{\dot{A}} &=& \frac12(-\Ga^{\ast})^m\Ga^{K{L}}_{\dot{A}\dot{B}}{\chi}^{\dot{B}},\nn
\\
(S^{(m)B}{\psicompt_2})^I &=& 0,\nonumber\\
(S^{(m)B}{\psicompt})^I &=& m(-\Ga^{\ast})^m\Ga^{I}_{B\dot{B}}\chi^{\dot{B}},\nonumber\\
(S^{(m)B}{\chi})^{\dot{A}} &=& m(-\Ga^{\ast})^m\Ga^{I}_{B\dot{A}}{\psicompt_2}^{I}. \label{omskrivet2}
\ee
Using instead the definition (\ref{gammastern}) of $\Ga^*$ as the $(32
\times 32)$ matrix $\Ga^1\Ga^0$ means that we consider the $SO(16)$
vectors as $SO(9)$ spinors, and the $SO(16)$ spinor $\chi^{\dot{A}}$ 
as eight vector components of a $SO(9)$ vector-spinor. We can thus
relate them to the  gravitino
in section~\ref{vssec}. This is done by splitting the vector, spinor
and conjugate spinor indices of $SO(16)$ into those of $SO(8)$, 
and relating the corresponding gamma matrices to each other, as
described in appendix \ref{appA}. 
In appendix \ref{appB}, finally, we explain how to express the
generators (\ref{ke9kompakt}) of $\KEn$ in the basis
$(S^{(m)IJ},\,A^{(m)A})$. 
We can then act with the generators (\ref{ke9kompakt}) on the fields
$(\chi^{\dot{A}},\,\psicompt^I,\,\psicompt_2^I)$ according to
(\ref{omskrivet2}) 
and require that the result, expressed in ($\psicomp_\al,\,\eta$),
coincide with the transformations of these expressions under $\KEn$
according to 
(\ref{ke9trmalfull}) and (\ref{ke9trm1full}). It turns out that this
requirement uniquely fixes the correspondance (\ref{lsferm}), in
agreement with the dimensional reduction \cite{Nicolai:1986jk}, up to
a constant factor multiplying all fields, and an arbitrary multiple of
$\psicompt^I_2$ that can be added to $\psicompt^I$.
In this fashion, we have recovered precisely the results of \cite{NiSa05}.

\end{subsection}

\end{section}

\vspace{3mm}

\noindent
{\bf{Acknowledgments:}} We are very grateful to T. Damour for numerous
enlightening discussions.

\appendix

\begin{section}{Gamma matrix conventions}\label{appA}

In this appendix, and the following one, we will no longer
follow the index convention for $\alpha,\,\beta,\ldots$,
introduced in section \ref{e9sec}.
Instead we will use $\alpha$ and $\dot{\alpha}$ as $SO(8)$ spinor
and conjugate spinor indices, respectively, while the indices $i,\,j,\ldots$
still take the values $3,\,\ldots,\,10$ as $SO(8)$ vector indices.
The chiral $(8\times 8)$ $SO(8)$ gamma-matrices will be denoted by
$\ga^i_{\al\db}$.

Then eight real, symmetric $(16 \times 16)$ gamma
matrices of $SO(9)$ can  be written
\be \label{so9gamma}
\ga^i_{IJ} =
\left(\begin{array}{cc}0&\ga^i_{\al\dot{\beta}}\\\ga^i_{\da\beta}&0\end{array}\right)\,, 
\ee
where $\ga^i_{\da\beta}$ is the transpose of $\ga^i_{\al\dot{\beta}}$.
The first eight $SO(9)$ gamma matrices square to one, anticommute, and
define the ninth matrix by
\be
\ga^3\cdots \ga^{10} =
\left(\begin{array}{cc}1&0\\0&-1\end{array}\right) \equiv \ga^{2}.
\ee
Thus $\ga^2$ also squares to one, and anticommutes with $\ga^i$.
The $SO(9)$ gamma matrices can be
  extended to the ten, real, symmetric $(32\times 32)$ gamma matrices
  of $SO(10)$, introduced in section \ref{dssec},  via
  \be
\Ga^{1} = \left(\begin{array}{cc}0&1\\1&0\end{array}\right),
\quad\quad \Ga^2 = \left(\begin{array}{cc}\ga^2&0\\0&-\ga^2\end{array}\right), \quad\quad
  \Ga^i = \left(\begin{array}{cc}\ga^i&0\\0&-\ga^i\end{array}\right).
\ee
In these conventions, the decomposition under $\Ga^2,\,\Ga^i$ of a 32 component spinor into two chiral spinors is manifest.
The $SO(10)$ gamma matrices satisfy
\be
\Ga^1\cdots \Ga^{10}=\left(\begin{array}{cc}0&-1\\1&0\end{array}\right)\equiv\Ga^{0}
\ee
and then we get
\be
\Ga^* \equiv \Ga^1\Ga^0 = \left(\begin{array}{cc}1&0\\0&-1\end{array}\right).
\ee

Triality implies that the matrices $\ga^\al_{i \dot{\beta}}$ and $\ga^{\dot{\beta}}_{i \al}$ have the same properties as $\ga^i_{\da\beta}$, and can also be extended to $SO(9)$ matrices as in (\ref{so9gamma}).
Thus we can take as $SO(16)$ gamma matrices the tensor products
\be
\Ga^{\alpha} = \bone \otimes \ga^\al,\nn\\
\Ga^{\da} = \ga^\da \otimes \ga^2 ,
\ee
with the components
\begin{align} \label{so16gamma}
\Ga^\al_{\beta\dot{\gamma},\delta j} &= \delta_{\beta\delta} \ga^j_{\al\dot{\gamma}},&\quad
  \Ga^\al_{ij,k\dot{\delta}} &= \delta_{ik} \ga^j_{\al\dot{\delta}},&\nn\\
\Ga^{\da}_{ij,\delta k} & = \delta_{jk} \ga^i_{\delta\dot{\al}},&\quad
  \Ga^{\da}_{\beta\dot{\gamma},i \dot{\delta}} &= - \delta_{\dot{\gamma}\dot{\delta}} \ga^i_{\beta\dot{\al}},&
\end{align}
as in \cite{Nicolai:1986jk}, and all other components are zero.
From this one can compute the following non-trivial anti-symmetric products
$\Ga^{IJ}_{AB}$ of gamma matrices,
\begin{align}
\Ga^{\al\beta}_{ij,kl} &= \delta_{ik} \ga^{jl}_{\al\beta},&
  \Ga^{\al\beta}_{\gamma\dot{\al},\delta\dot{\beta}} &= \delta_{\gamma\delta} \ga^k_{\dot{\al}[\al}
  \ga^k_{\beta]\dot{\beta}},&\nn\\
\Ga^{\alpha\dot{\beta}}_{ij,\ga\dot{\delta}} &=
   -\ga^i_{\ga\dot{\beta}} \ga^j_{\al\dot{\delta}},&
  \Ga^{\al\dot{\beta}}_{\ga\dot{\delta},ij} &= 
    \ga^i_{\ga\dot{\beta}}\ga^j_{\alpha\dot{\delta}},&\nn\\
\Ga^{\dot{\al}\dot{\beta}}_{ij,kl} &= \delta_{jl} \ga^{ik}_{\dot{\al}\dot{\beta}},&
   \Ga^{\da\dot{\beta}}_{\al\dot{\gamma},\beta\dot{\delta}} &= \delta_{\dot{\ga}\dot{\delta}} \ga^k_{\dot{\al}[\al}
  \ga^k_{\beta]\dot{\beta}}.&
\end{align}
In (\ref{so16gamma}), we see that the vector, spinor and conjugate spinor indices $(I,\,A,\,\dot{A})$ of $SO(16)$ split into those of $SO(8)$ as
\be
I &=& (\al,\,\dot{\al}),\nn\\
A &=& (\al\dot{\al},\,ij),\nn\\
\dot{A} &=& (\al i,\, j \dot{\al}),
\ee
according to the decompositions
\begin{alignat}{2}
{\bf 16} \quad&\rightarrow& \quad {\bf ({8}_c,\overline{1})} {\bf \oplus (1,\overline{8}_s)}
  \quad\rightarrow& \quad {\bf 8_s}\oplus{\bf 8_c},\nn\\
{\bf 128_s} &\rightarrow& \quad{\bf (8_v,\overline{8}_v)}  {\bf \oplus (8_s,\overline{8}_c)}
  \quad\rightarrow& \quad {\bf 1}\oplus{\bf 28}\oplus{\bf 35_v}\oplus{\bf 8_v}\oplus{\bf 56_v},\nn\\
{\bf 128_c} &\rightarrow& \quad{\bf (8_v,\overline{8}_c)}  {\bf \oplus (8_s,\overline{8}_v)}
  \quad\rightarrow& \quad {\bf 8_s \oplus 56_s \oplus 8_c \oplus 56_c} \label{A12}
\end{alignat}
of these $\mf{so}(16)$ representations under $\mf{so}(8) \oplus \mf{so}(8)$, and then under the diagonal $\mf{so}(8)$ subalgebra.
For example, the first line in (\ref{lsferm}) then reads
\be\label{so16covspinors}
\chi^{i\dot{\al}}_\pm &=&  (\psicomp^i_\pm)^{\dot{\al}} -
  \frac12(\ga^i\ga^j)_{\dot{\al}\dot{\beta}}({\psicomp^{j}_\pm})^{\dot{\beta}},\nn\\
\chi^{{\al}i}_\pm &=&  (\psicomp^i_\pm)^{{\al}} -
  \frac12(\ga^i\ga^j)_{\al\beta}({\psicomp^{j}_\pm})^{\beta}.
\ee

\end{section}

\begin{section}{Relation between the two $E_8$ bases}\label{appB}

We have in this article used two different bases of $E_{8}$. The first one arose in the $A_7$ level decomposition described in section \ref{e8viaa7} (table \ref{tabe8a7}), and for the compact generators it was generalized to $E_9$ in section \ref{ke9sec}. The second one, covariant under the maximal compact subalgebra $\mf{so}(16)$, was introduced in section \ref{ke9curr}, and extended to $E_9$ via the current algebra construction. We will now explain the relation between these two bases, which was also given in \cite{KoNiSa00} but in different
conventions.
First, as for $E_9$ in section \ref{ke9sec}, we consider the compact linear combinations
\be \label{kompakta}
J^{ij} &=& G^i_j-G^j_i,\nn\\
J^{ijk} &=& Z^{ijk}-Z_{ijk},\nn\\
J^{i_1 \dots i_6} &=& Z^{i_1 \dots i_6}-Z_{i_1 \dots i_6},\nn\\
J^i &=& Z^i - Z_i
\ee
of the basis elements in table \ref{tabe8a7}.
These can now be expressed in $X^{IJ}$ by the $SO(9)$ or $SO(8)$ gamma matrices as
\begin{alignat}{2}
J^{ij} &= \frac14 \ga^{ij}_{IJ} X^{IJ} &&=\frac14 \ga^{ij}_{\al\beta} X^{\al\beta}+\frac14 \ga^{ij}_{\da\dot{\beta}} X^{\da\dot{\beta}},\nn\\
J^{i_1i_2i_3} &= -\frac14  \ga^{i_1i_2i_3}_{IJ} X^{IJ} &&=-\frac12 \ga^{i_1i_2i_3}_{\al\dot{\beta}} X^{\al\dot{\beta}},\nn\\
J^{i_1\ldots i_6} &= \frac14 \ga^{i_1\ldots i_6}_{IJ} X^{IJ} &&=
\frac14 \ga^{i_1\ldots i_6}_{\al\beta} X^{\al\beta} + \frac14 \ga^{i_1\ldots i_6}_{\da\dot{\beta}} X^{\da\dot{\beta}},\nn\\
J^i &= -\frac14  (\ga^i\ga^2)_{IJ} X^{IJ} &&= -\frac12  (\ga^i\ga^2)_{\al\dot{\beta}} X^{\al\dot{\beta}}, \label{JX}
\end{alignat}
where a sign ambiguity in the derivation has been fixed
by demanding that the generalised evaluation map
(\ref{lstrm}) and the representation given by (\ref{ke9trmalfull})
and (\ref{ke9trm1full}) agree.
For the remaining generators
\be\label{sgens}
S^{ij} &=& G^i{}_j + G^j{}_i,\nn\\
S^{i_1i_2i_3} &=& Z^{i_1i_2i_3} + Z_{i_1i_2i_3},\nn\\
S^{i_1\ldots i_6} &=& Z^{i_1\ldots i_6} + Z_{i_1\ldots i_6},\nn\\
S^i &=& Z^i + Z_i,
\ee
it is necessary to break $SO(9)$ covariance, and split the $SO(16)$ spinor indices. Then we get
\be\label{SY}
S^{ij} &=& 2 Y^{(ij)} - \delta^{ij} Y^{kk},\nn\\
S^{i_1i_2i_3} &=& -
  \frac12\ga^{i_1i_2i_3}_{\al\dot{\beta}}Y^{\al\dot{\beta}},\nn\\
S^{i_1\ldots i_6} &=&  \eps^{i_1\ldots i_6k_1k_2}
  Y^{k_1k_2},\nn\\
S^i &=& -\frac12 \ga^i_{\al\dot{\beta}} Y^{\al\dot{\beta}},
\ee
where an overall sign ambiguity in the definition of $Y^A$ has been fixed
again by equivalence between the representations.
Combining these formulas with (\ref{aff1}), (\ref{affm1}), (\ref{e9e10a}) and (\ref{e9e10b}), we 
can easily express the generators (\ref{ke9kompakt}) in the basis $(S^{(m)IJ},\,A^{(m)A})$ introduced in section
\ref{soapp}. For example, we have
\be \label{basrelation}
J^{ijk}_{(3m+1)}&=&- \ga^{ijk}_{\al \dot{\beta}}S^{(m)\al\dot{\beta}}- \ga^{ijk}_{\al \dot{\beta}}A^{(m)\al\dot{\beta}},\nn\\
\frac1{5!}\epsilon^{ijk l_1\dots l_5}J^{l_1\dots l_5 2}_{(3m+2)}
&=&\ga^{ijk}_{\al \dot{\beta}}S^{(m)\al\dot{\beta}}- \ga^{ijk}_{\al \dot{\beta}}A^{(m)\al\dot{\beta}},
\ee
and in the same way we obtain the remaining, non-compact,
generators of $E_9$.

\end{section}


\begin{thebibliography}{30}

\bibitem{NiSa05} H.~Nicolai and H.~Samtleben, {\sl On $K(E_9)$},
  Q.\ J.\ Pure Appl.\ Math.\  {\bf 1} (2005) 180, {\tt hep-th/0407055}

\bibitem{DaHeNi02} T.~Damour, M.~Henneaux and H.~Nicolai, {\sl
    $E_{10}$ and a "small tension expansion" of M-theory}, Phys.
    Rev. Lett. {\bf 89} (2002) 221601, {\tt hep-th/0207267}

\bibitem{DaNi04} T.~Damour and H.~Nicolai, {\sl Eleven dimensional
  supergravity and the $E_{10}/$ $K(E_{10})$ $\sigma$-model at low $A_9$
  levels}, in: Group Theoretical Methods in Physics, Institute of
  Physics Conference Series No. 185, IoP Publishing, 2005,
  {\tt hep-th/0410245}

\bibitem{We00} P.~C.~West, {\sl Hidden superconformal symmetry in
    M theory}, JHEP {\bf 0008} (2000) 007, {\tt hep-th/0005270}

\bibitem{We01} P.~C.~West, {\sl $E_{11}$ and M theory}, Class.
    Quant. Grav. {\bf 18} (2001) 4443--4460, {\tt hep-th/0104081}

\bibitem{EnHo04} F.~Englert and L.~Houart, {\sl $\cal{G}^{+++}$
    invariant formulation of gravity and M-theories: exact BPS
    solutions}, JHEP {\bf 0401} (2004) 002, {\tt hep-th/0311255}

\bibitem{KlNi05} A.~Kleinschmidt and H.~Nicolai, {\sl Gradient
  representations and affine structures in $AE_n$}, Class. Quantum
  Grav. {\bf 22} (2005) 4457--4488, {\tt hep-th/0506238}

\bibitem{dBHP05} S.~de Buyl, M.~Henneaux and L.~Paulot, {\sl Hidden
  symmetries and Dirac fermions},  Class.\ Quant.\ Grav.\  {\bf 22}
  (2005) 3595, {\tt hep-th/0506009}

\bibitem{DaKlNi06a} T.~Damour, A.~Kleinschmidt and H.~Nicolai, {\sl
  Hidden symmetries and the fermionic sector of eleven-dimensional
  supergravity},   Phys.\ Lett.\ B {\bf 634} (2006) 319, {\tt
  hep-th/0512163}

\bibitem{dBHP06} S.~de Buyl, M.~Henneaux and L.~Paulot, {\sl Extended
  $E_8$ invariance of 11-dimensional supergravity},  JHEP {\bf 0602}
  (2006) 056, {\tt hep-th/0512292}

\bibitem{DaKlNi06b} T.~Damour, A.~Kleinschmidt and H.~Nicolai, {\sl
  $K(E_{10})$, supergravity and fermions}, JHEP {\bf 08} (2006) 048,
  {\tt hep-th/0606105}

\bibitem{KlNi06} A.~Kleinschmidt and H.~Nicolai, {\sl IIA and IIB
  spinors from $K(E_{10})$}, Phys. Lett. B {\bf 637} (2006) 107--112,
  {\tt hep-th/0603205}

\bibitem{KlNi04a} A.~Kleinschmidt and H.~Nicolai, {\sl $E_{10}$ and
  $SO(9,9)$ invariant supergravity}, JHEP {\bf 0407} (2004) 041, {\tt
  hep-th/0407101}

\bibitem{KlNi04b}  A.~Kleinschmidt and H.~Nicolai, {\sl IIB
  supergravity and $E_{10}$}, Phys.\ Lett.\ B {\bf 606} (2005) 391,
  {\tt hep-th/0411225}

\bibitem{SchnWe02} I.~Schnakenburg and P.~C.~West, {\sl Massive
    IIA supergravity as a nonlinear realization}, Phys. Lett. B
    {\bf 540} (2002) 137--145, {\tt hep-th/0204207}

\bibitem{SchnWe01} I.~Schnakenburg and P.~C.~West, {\sl Kac--Moody
    symmetries of IIB supergravity}, Phys. Lett. B {\bf 517} (2001)
    421--428, {\tt hep-th/0107081}

\bibitem{We04a} P.C.~West, {\sl The ${\rm IIA}$, ${\rm IIB}$ and
    eleven-dimensional theory and their common $E_{11}$ origin},
    Nucl. Phys. B {\bf  693} (2004) 76--102,
    {\tt hep-th/0402140}

\bibitem{HiKl06} C.~Hillmann and A.~Kleinschmidt, {\sl Pure type~I
  supergravity and $DE_{10}$}, Gen. Rel. Grav. {\bf 38} (2006)
  1861--1885, {\tt hep-th/0608092}

\bibitem{SchnWe04} I.~Schnakenburg and P.~West, {\sl Kac-Moody
  Symmetries of Ten-dimensional Non-maximal Supergravity
  Theories}, JHEP {\bf 0405} (2004) 019, {\tt hep-th/0401196}

\bibitem{Ju80} B.~Julia,
  {\sl Kac--Moody Symmetry of Gravitation
  and Supergravity Theories}, in: M.~Flato, P.~Sally
  and G.~Zuckerman (eds.), Applications of Group Theory in Physics
  and Mathematical Physics (Lectures in Applied Mathematics {\bf
  21}), Am. Math. Soc. (Providence, 1985) 355--374, LPTENS 82/22

\bibitem{Ni87}  H.~Nicolai, {\sl The integrability of $N=16$ supergravity},
  Phys.\ Lett.\ B {\bf 194} (1987) 402

\bibitem{NiWa89}  H.~Nicolai and N.~P.~Warner, {\sl The structure of
  $N=16$ supergravity in two dimensions}, Commun.\ Math.\ Phys.\  {\bf
  125} (1989) 369

\bibitem{NiSa98}   H.~Nicolai and H.~Samtleben, {\sl Integrability and
  canonical structure of $d = 2, N = 16$ supergravity},
  Nucl.\ Phys.\ B {\bf 533} (1998) 210, {\tt hep-th/9804152}

\bibitem{Ge71} R.~Geroch, {\sl A method for generating solutions
  of Einstein's equations}, J. Math. Phys. {\bf 12} (1971)
  918--924; {\sl A method for generating solutions
  of Einstein's equations. II}, J. Math. Phys. {\bf 13} (1972)
  394--404

\bibitem{BrMa87} P.~Breitenlohner and D.~Maison, {\sl On the
    Geroch group}, Ann. Poincar\'e Phys. Theor. {\bf 46} (1987)
    215--246

\bibitem{CrJu78}  E.~Cremmer and B.~Julia, {\sl The $SO(8)$
  Supergravity},   Nucl.\ Phys.\ B {\bf 159} (1979) 141

\bibitem{CrJuLuPo98}  E.~Cremmer, B.~Julia, H.~Lu and C.~N.~Pope,
  {\sl Dualisation of dualities. I},
  Nucl.\ Phys.\ B {\bf 523} (1998) 73, {\tt hep-th/9710119}

\bibitem{FiNi03} H. Nicolai and T. Fischbacher, { Low Level
  representations of $E_{10}$ and $E_{11}$}, Contribution to the
  Proceedings of
  the Ramanujan International Symposium on Kac-Moody Algebras and
  Applications, ISKMAA-2002, Jan. 28--31, Chennai, India, {\tt
  hep-th/0301017}

\bibitem{We03a}   P.C.~West, {\sl Very extended $E_8$ and $A_8$ at low
  levels, gravity and supergravity},
  Class.\ Quant.\ Grav.\  {\bf 20} (2003) 2393, {\tt hep-th/0212291}

\bibitem{DuLiu03} M.~Duff and J.~T.~Liu, {\sl Hidden space-time
  symmetries and generalized holonomy in M theory}, Nucl. Phys. B
  {\bf 674} (2003) 217--230, {\tt hep-th/0303140}

\bibitem{Hu03} C.~Hull, {\sl Holonomy and symmetry in M theory},
  {\tt hep-th/0305039}

\bibitem{DuSt91} M.~J.~Duff and K.~S.~Stelle, {\sl Multimembrane
  solutions of $D=11$ supergravity}, Phys. Lett. B {\bf 253}
  (1991) 113--118

\bibitem{BaDuLiWe05}  A.~Batrachenko, M.~J.~Duff, J.~T.~Liu and
  W.~Y.~Wen, {\sl Generalized holonomy of M-theory vacua}, Nucl.\
  Phys.\ B {\bf 726} (2005) 275, {\tt hep-th/0312165}

\bibitem{Ke04} A.~Keurentjes, {\sl The topology of U Duality
  (sub)groups}, Class. Quantum. Grav. {\bf 21} (2004) 1695-1708,
  {\tt hep-th/0309106}

\bibitem{KoNiSa00}  K.~Koepsell, H.~Nicolai and H.~Samtleben, {\sl An
  exceptional geometry for $d = 11$ supergravity?},
 Class.\ Quant.\ Grav.\  {\bf 17} (2000) 3689, {\tt hep-th/0006034}

\bibitem{Ka90} V.~G.~Kac, {\sl Infinite dimensional Lie algebras}, 3rd
  edition, Cambridge University Press (Cambridge, 1990)

\bibitem{Nicolai:1986jk}   H.~Nicolai, {\sl $D = 11$ supergravity with local
  $SO(16)$ invariance},   Phys.\ Lett.\  B {\bf 187} (1987) 316

\end{thebibliography}
\end{document}